\documentclass[onecolumn]{aastex631}

\shorttitle{Modelling Non-Condensing Compositional Convection}
\shortauthors{Habib \& Pierrehumbert 2022}

\usepackage{float}
\usepackage{booktabs}
\usepackage{mathtools}

\begin{document}
\title{Modelling Non-Condensing Compositional Convection for Applications to super-Earth and sub-Neptune Atmospheres}

\author[0000-0001-7131-251X]{Namrah Habib}
\affiliation{Atmospheric, Oceanic and Planetary Physics, Department of Physics, University of Oxford, Parks Road, OX1 3PU, Oxford, UK}

\author[0000-0002-5887-1197]{Raymond T. Pierrehumbert}
\affiliation{Atmospheric, Oceanic and Planetary Physics, Department of Physics, University of Oxford, Parks Road, OX1 3PU, Oxford, UK}

\begin{abstract}

Compositional convection is atmospheric mixing driven by density variations caused by compositional gradients. Previous studies have suggested that compositional gradients of atmospheric trace species within planetary atmospheres can impact convection and the final atmospheric temperature profile. In this work, we employ 3D convection resolving simulations using Cloud Model 1 (CM1) to gain a fundamental understanding of how compositional variation influences convection and the final atmospheric state of exoplanet atmospheres. We perform 3D initial value problem simulations of non-condensing compositional convection for Earth-Air, $\rm H_2$, and $\rm CO_2$ atmospheres. Conventionally, atmospheric convection is assumed to mix the atmosphere to a final, marginally stable state defined by a unique temperature profile. However, when there is compositional variation within an atmosphere, a continuous family of stable end states is possible, differing in the final state composition profile. Our CM1 simulations are used to determine which of the family of possible compositional end states is selected. Leveraging the results from our CM1 simulations, we develop a dry convective adjustment scheme for use in General Circulation Models (GCMs). This scheme relies on an energy analysis to determine the final adjusted atmospheric state. Our convection scheme produces results that agree with our CM1 simulations and can easily be implemented in GCMs to improve modelling of compositional convection in exoplanet atmospheres.

\end{abstract}

\section{Introduction} \label{sec:intro}

3D climate models, or General Circulation Models (GCMs), are one of the key tools used to study and understand the climate of exoplanets. Representation of convection within GCMs is a known source of uncertainty \citep{Fauchez2020THAI, Yang2019, Fauchez2019, Sergeev2020} in climate modelling of exoplanet atmospheres, especially for smaller ~$\rm 1-10~R_{Earth}$ sized planets. GCMs resolve planetary atmospheric circulations on the order of hundreds of km, whereas convection can occur on the order of meter to kilometer length scales \citep{Atkinson1996mesoscale}. Therefore, convection is parameterized in GCMs. There are two main types of convection parameterizations, convective adjustment schemes and mass flux parameterizations.

Convective adjustment schemes adjust the vertical atmospheric sounding state to a marginally stable state while conserving enthalpy. In most cases, the marginally stable state in convective adjustment schemes is defined to be a moist or dry adiabat. Alternatively, mass-flux parameterizations simulate the statistical effect of convective plumes without resolving convection \citep{Arakawa2004}. Mass flux schemes are considered to be more representative of convection, however, they are based on years of Earth convection data and are often ``tuned" to model present-day Earth. \cite{Sergeev2020} compared GCM outputs for two terrestrial exoplanets, Proxima-b and TRAPPIST-1e, using the two different types of convection schemes and a high resolution convection resolving model. Their study showed that convective adjustment schemes predicted higher atmospheric temperatures and enhanced zonal flows compared to mass flux schemes. However, mass-flux convection schemes cannot be taken as ground-truth for modelling exoplanet atmospheres since they contain free parameters that are difficult to tune without access to detailed observations such as those available for Earth. Additionally, \cite{Sergeev2020} performed simulations which used a convection resolving model in a nested grid configuration with their GCM. Their study suggests that convection parameterization schemes overestimate the average surface temperature pattern relative to convection permitting simulations.

Recently, the TRAPPIST-1 Habitable Atmosphere Intercomparison (THAI) Protocol \citep{Fauchez2020THAI, Turbet2021THAI, Sergeev2021THAI, Fauchez2021THAI} compared 3D climate modelling output from four GCMs (ExoCAM, LMD-Generic, ROCKE-3D, and UM) that are commonly used to model rocky exoplanets. The THAI study showed that all four GCMs predicted similar climate but differ in the details of the climate regime \citep{Turbet2021THAI, Sergeev2021THAI}. In particular, \cite{Sergeev2021THAI} showed that cloud and convection parameterizations are one of the major sources of the differences within the 3D simulated climate output from the four GCMs. The THAI study illustrates the need to improve our understanding and modelling of atmospheric convection in GCMs so we can better represent and study the climate of exoplanet atmospheres.

We consider modelling atmospheric convection for applications to a diverse range of super-Earth and sub-Neptune planetary atmospheres, including compositions that differ greatly from Earth's current atmosphere. Super-Earth and sub-Neptune planets are approximately between the size of Earth and Neptune ($\rm ~1.2~-~3.9~R_{Earth}$), and are the most commonly detected exoplanet from the \textit{Kepler Mission} but have no true analog in the Solar System. In this work, we make the distinction between super-Earths and sub-Neptunes based on the planet's density. Super-Earths are planets with a density roughly consistent with a rocky composition, and are typically thought to have secondary, higher mean molecular weight atmospheres over a rocky core \citep{Seager2007}. Sub-Neptunes are low-density planets thought to have low density volatile envelopes making up a substantial proportion of the mass of the planet \citep{Rogers2011, VanEylen2018}. A commonly accepted model for sub-Neptune structure, preferred by observational constraints \citep{VanEylen2018}, is a planet with a hydrogen helium dominated atmosphere over a rocky or icy core. Other proposed sub-Neptune structure includes ``Hycean Worlds" \citep{Madhusudhan2021}, which are planets with a hydrogen helium dominated atmosphere with a water rich interior and large oceans. In this work we will use the term sub-Neptune to refer to exoplanets that are $\rm 1.2~-~3.9~R_{Earth}$ and have a substantial fraction of hydrogen within the atmosphere. We consider a range of $\rm H_2-H_2O$ mixing ratios to represent the current thinking about the compositional space of sub-Neptune atmospheres.

Convection parameterizations for exoplanet atmospheres are challenged by the greater diversity of exoplanet atmospheric compositions compared to Earth. For many sub-Neptune and super-Earth atmospheres, the effects of vertically inhomogeneous atmospheric trace gases on convection are important \citep{Guillot1995, Leconte2017, Li2015, Markham2022}. Compositional convection is a convective process in which buoyancy is significantly influenced by the compositional gradient of atmospheric tracers. In Earth’s present atmosphere, $\rm H_2O$ has a lower molecular weight compared to $\rm N_{2}$ air and contributes slightly to the positive buoyancy of an air parcel. The compositional convection effect in Earth’s atmosphere is small, but has been shown to significantly affect the Earth's climate in the tropics \citep[see][]{Seidel2020, Yang2020}. In higher mean molecular weight atmospheres, such as $\rm CO_2$ dominated atmospheres, water vapor further enhances the positive buoyancy of an atmospheric air parcel leading to a larger compositional buoyancy effect. \cite{DaleyYates2021} used 3D hydrodynamics simulations to study compositional convection for hot, rocky exoplanets with a chemical reaction forcing term. Their simulations showed that compositional convection leads to a reduction of the temperature gradient, causing the lower atmosphere temperature to decrease while the upper atmosphere temperature increased.

Conversely, in sub-Neptune exoplanets with primarily ${\rm H_{2}/He}$ atmospheres, convection can be inhibited by the greater density of atmospheric tracers compared to the background air \citep{Guillot1995, Leconte2017}. When accounting for compositional gradients of water vapour and other atmospheric tracers in the $\rm H_2 /He$ atmospheres of the Solar System giant planets, \cite{Leconte2017} showed that the planetary deep atmospheric temperature can be hundreds K hotter than predicted by a moist or dry adiabat. Hotter deep atmospheric temperatures impact our understanding of planetary interior chemistry, evolution, and habitability for terrestrial exoplanets. Previous studies of convective inhibition due to compositional gradients have focused on modelling the Solar System giant planets \citep{Guillot1995, Leconte2017, Li2015}. However, compositional convection within $\rm H_2$/He atmospheres is relevant for sub-Neptunes atmospheres, terrestrial planets with a persistent $\rm H_2$ supported greenhouse habitability such as those discussed in \cite{Pierrehumbert2011}, and transient primordial $\rm H_2$ atmospheres that have accreted on rocky planets while the protoplanetary disk is still present \citep{Young2023earth}.

We perform 3D convection resolving simulations of an atmospheric system with an initial compositional gradient that creates a high density atmospheric layer above a lower density atmospheric layer. A layer of high density fluid situated above a layer of lower density fluid is unstable and will ultimately lead to turbulent mixing that eliminates the instability. This is known as the Rayleigh-Taylor instability, and constitutes one of the classic problems of fluid mechanics \citep{bud1992stabilization, bernstein1983effect, Gauthier2010}. Our work extends the study of turbulent mixing by the Rayleigh-Taylor instability to a wide range of situations relevant to planetary atmospheres, incorporating the effects of composition on buoyancy.

Specifically, we perform 3D nonlinear, initial value problem simulations of initially unstable atmospheric profiles involving mixtures of water vapor with Earth air, $\rm CO_2$, and $\rm H_2$ atmospheres using the 3D non-hydrostatic, convection resolving model, Cloud Model 1 \citep[CM1,][]{Bryan2009}. We explore both compositional suppression of buoyancy, for applications to $\rm H_2$ atmospheres, and compositional enhancement of buoyancy, for applications to higher mean molecular weight atmospheres. In an unforced initial value problem, an initial perturbation to the state of rest triggers the conversion of potential energy into kinetic energy, mixing the system and eventually, viciously dissipating as heat. Mixing results in a new state of rest which must be stable to at least infinitesimal perturbations. It is commonly assumed that a displaced parcel in the end state has zero buoyancy relative to the surroundings, so that it neither rises nor sinks. Such states are defined as ``marginally stable." In essence, convection is assumed to do the minimum amount of mixing needed to prevent further growth of any perturbations to the end state.

Our study aims to determine how convective mixing in an initial, unstable inhomogeneous composition atmosphere affects the final atmospheric temperature and compositional profiles. With compositional variations, density depends on both composition and temperature. Therefore, there is a continuous family of marginally stable states that the system could mix to, differing in the final compositional profile (see Section \ref{sec:atm_thermo_background}). A chief goal of our work is to shed light on the which of the possible final compositional profiles is left behind after convective mixing.

Further, we use our initial value problem simulations to inform the development of a convective adjustment scheme. Any convective adjustment scheme must; (1.) determine if the atmosphere adjusts to a marginally stable state, (2.) identify the temperature and compositional profiles that represent this marginally stable state. For a homogeneous composition atmosphere, convection is assumed to mix to a marginally stable state that follows an adiabat and uniquely defines the end state. However, when the initial state has a compositional gradient, assuming convection mixes to a final marginally stable state no longer uniquely defines the end state.

In a realistic atmospheric situation, convective instability would be continually created by radiative heating and cooling or by heat flux from the planetary interior. The rate of generation of instability is balanced against the rate of relaxation towards a marginally stable profile by buoyancy-driven fluid mixing, leading to a statistical-equilibrium turbulent state. The statistical equilibrium problem is more challenging to analyze, but initial value problems provide essential information about the pace of adjustment and the layer affected by mixing. If the turbulent mixing is rapid compared to the rate of regeneration of instability, then the initial value problem provides direct information about the equilibrium state. This is the underlying assumption of convective adjustment parameterizations.

In the present work, we restrict our attention to situations in which water vapor does not condense, in order to explore 3D compositional convection in its simplest form. Our simulation results apply to exoplanets which have atmospheres that are too hot for water to condense within the convecting atmospheric layer. Examples of such exoplanets include; sub-Neptune GJ~1214b, if we assume a $\rm H_2$ dominated atmosphere, sub-Neptune K2-18b where $\rm H_2O$ condensation is limited to a thin layer of the upper atmosphere, and super-Earth 55 Cnc e, if we assume a CO or $\rm N_2$ based atmosphere. The work presented here paves the way for future CM1 simulations to explore compositional convection with condensation. We use the results from our non-condensing CM1 simulations to inform the development of a convective adjustment scheme that can better represent non-condensing compositional convection in GCM modelling of super-Earth and sub-Neptune atmospheres, which is a necessary first step as any convection scheme must be able to handle the non-condensing case accurately.

In section~\ref{sec:atm_thermo_background} we review atmospheric thermodynamics concepts that are relevant for this work and present the definition of the virtual adiabat. In section~\ref{sec:model_description} we describe our CM1 initial value problem simulations, and present our simulation results in section~\ref{sec:cm1_results}. We use our CM1 results to develop a simple convective adjustment scheme which is discussed in section~\ref{sec:adj_scheme}.

\section{Atmospheric Thermodynamics} \label{sec:atm_thermo_background}

Assessment of the stability of an initial atmospheric state can proceed from a linearized stability analysis or a buoyancy analysis in which the density of a displaced parcel is compared to the density of the surrounding atmosphere. The former kind of analysis yields how initial instability growth rates depend on the horizontal wavelength of a perturbation. Alternatively, buoyancy analysis, which is more common in the context of atmospheric convection studies, has the advantage that it considers the energy available for mixing when air parcels are displaced by a finite amount, and this is the approach employed here. As a basis for convective adjustment, which assumes that the final adjusted state is marginally stable, it is essential to be able to determine the stability properties of a given profile. The best form of stability that can be obtained is neutral stability, in which an initial displacement doesn't grow in amplitude in the absence of dissipation. A displaced parcel in a neutrally stable situation will experience buoyancy that acts to move it back towards its original position, but without dissipation the parcel will oscillate about its original position without decay. Although, it is expected that friction will eventually damp out the oscillations. We can further distinguish local versus global neutral stability. Local stability only demands neutral stability for infinitesimal displacements, whereas global stability demands neutral stability against displacements with a finite (though possibly bounded) amplitude. Marginally stable states are neutrally stable states where a displaced parcel has zero buoyancy, and therefore remains in place rather than oscillating about its original position.

Stability can be determined by adiabatically displacing an atmospheric parcel vertically, without allowing any mixing of the parcel with its surroundings, and comparing its density with the ambient atmosphere at its new level. We will consider mixtures of a background gas, with corresponding quantities labeled with subscript $b$, with a second gas denoted by subscript $v$ (for ``vapor"). Since we don't consider condensation and the second gas is not restricted to be dilute in the background, the designation of which gas to consider the background is arbitrary. The notation is chosen to smooth the path to future work in which the second gas is condensible and the background does not condense. In all the examples we consider explicitly in the current work, the second gas is water vapor. We work in a temperature and pressure range in which water vapor does not condense.

Let $\rm q_v$ be the mass mixing ratio of the vapor to the background gas, i.e. $\rm \rho_v/\rho_b$, where $\rm \rho$ is density. Up to a constant, the entropy per unit mass of the mixture is
\begin{equation}\label{eq:entropy}
    \rm S(p) = C_{p, mix} \ln{T} - R_b \left( \frac{1}{1 + q_v} \right) \ln{p_b} - R_v \left(\frac{q_v}{1 + q_v}\right) \ln{p_v},
\end{equation}
where $\rm p_b$ and $\rm p_v$ represent the partial pressure of the background and tracer components respectively \citep[][Chap.4]{Emanuel1994}. $\rm C_{p, mix}$ is the mass-weighted mean specific heat of the mixture, at constant pressure,
\begin{equation}
 \rm C_{p,\,mix}~=~\frac{1}{1 + q_v}C_{p,\,b} + \frac{q_v}{1 + q_v} C_{p,\,v}.
\end{equation}
The specific heats of the individual gases depend on temperature, but the temperature dependence will be neglected in the current work. However, $\rm C_{p,\,mix}$ still varies because of vertical variations in $\rm q_v$. $\rm R_b\equiv R^*/M_b$ and $\rm R_v\equiv R^*/M_v$ are the gas constants for the two gases, where $\rm R^*$ is the universal gas constant, and $\rm M_b$ and $\rm M_v$ are the molar masses of the respective gases. The total pressure is $\rm p = p_b+p_v$, and serves as a vertical coordinate.

Suppose a parcel with original temperature $\rm T(p_1)$ is displaced adiabatically to a pressure level $\rm p_2$, without any mixing with the surroundings. Using conservation of entropy, $\rm T(p_2)$ is determined by;
\begin{align}\label{eq:EntropyBalance}
0 &= C_{p, mix} \ln{\frac{T(p_2)}{T(p_1)}} - R_b \left( \frac{1}{1 + q_v} \right) \ln{\frac{p_{b,2}}{p_{b,1}}} - R_v \left(\frac{q_v}{1 + q_v}\right) \ln{\frac{p_{v,2}}{p_{v,1}}}\\
 &= C_{p, mix} \ln{\frac{T(p_2)}{T(p_1)}} - R_{mix} \ln{\frac{p_2}{p_1}}
\end{align}
where
\begin{equation}
    \rm R_{mix} = R_b \left( \frac{1}{1 + q_v} \right) + R_v \left(\frac{q_v}{1 + q_v}\right).
\end{equation}
Since the parcel is displaced without mixing, $\rm q_v$ is fixed at its original value, hence $\rm C_{p,mix}$ also remains fixed at its original value. Because $\rm q_v$ (and likewise molar mixing ratio) remains fixed, Dalton's Law implies $\rm p_{b,2}/p_{b,1} = p_2/p_1$ and $\rm p_{v,2}/p_{v,1} = p_2/p_1$, from which the second equality in Eqn.~\ref{eq:EntropyBalance} is derived. If we define $\rm \beta_{mix} = R_{mix}/C_{p,mix}$ then,
\begin{equation}\label{eq:DryAdiabat}
    \rm T(p_2) = T(p_1) \left( \frac{p_2}{p_1} \right) ^ {\beta_{mix}(p_1)}.
\end{equation}
Eqn.~\ref{eq:DryAdiabat} defines the dry adiabat originating from $\rm p_1$. Note that the logarithmic slope in Eqn.~\ref{eq:DryAdiabat} is given by the value of $\rm \beta_{mix}$ at the original pressure level $\rm p_1$.

To determine stability, we compare the density of the displaced parcel with the density of the ambient atmosphere at $\rm p_2$. The density of displaced and ambient parcels is determined from their temperature using the ideal gas equation of state with gas constants appropriate to their respective composition (i.e. mean molar mass). When composition is inhomogeneous, it is customary to compute density by introducing the virtual temperature, $\rm T_v$, defined such that $\rm p = \rho R_b T_v$. Thus,
\begin{equation} \label{eq:virtual_temp}
    \rm T_v = T \;\frac{M_b}{\mu(p)},
\end{equation}
where $\mu$ is the mass-weighted harmonic mean molar mass of the gas mixture and $\rm M_b$ is the molar mass of the background gas. The mean molar mass can be written as,
\begin{equation} \label{eq:MMW}
    \rm \frac{M_b}{\mu(p)} = \left( \frac{1+ {q_v (p)}/{\epsilon}}{1 + q_v (p)} \right),
\end{equation}
where $\rm \epsilon = {M_{v}}/M_b$, and $\rm M_v$ is the molar mass of the tracer. When $\rm \epsilon <1 $ the tracer enhances buoyancy, whereas for $\rm \epsilon >1 $ (e.g. for $\mathrm{H_2O}$ tracer in an $\mathrm{H_2}$ atmosphere) the tracer suppresses buoyancy.

For a parcel displaced adiabatically without mixing, the original value $\rm q_v(p_1)$ is conserved, so this value is used in computing the virtual temperature of the displaced parcel. Likewise, the original $\rm \beta_{mix}(p_1)$ is used in computing the temperature of the displaced parcel. The resulting virtual temperature of the adiabatically displaced parcel defines the virtual adiabat which passes through $\rm (p_1,T_v(p_1))$, for which we will use the notation $\rm T_{v,ad}(p_2|p_1)$. Thus,
\begin{equation}\label{eq:VirtualAdiabatExact}
    \rm T_{v,ad}(p_2|p_1) = T_v(p_1) \left( \frac{p_2}{p_1} \right) ^ {\beta_{mix}(p_1)}.
\end{equation}
The actual temperature of the displaced parcel, $\rm T_{ad}(p_2|p_1)$ can be obtained from Eqn.~\ref{eq:VirtualAdiabatExact} using Eqn.~\ref{eq:virtual_temp} and Eqn.~\ref{eq:MMW} with mixing ratio $\rm q_v(p_1)$ corresponding to that of the parcel at its origin.

A lifted parcel ($\rm p_2 < p_1$) has negative buoyancy, and hence is neutrally stable, when its virtual temperature $\rm T_{v,ad}(p_2|p_1)$ is less than the ambient virtual temperature $\rm T_{v,amb} (p_2)$. The lifted parcel's virtual temperature $\rm T_v(p_2)$ is computed from Eqn.~\ref{eq:VirtualAdiabatExact} using the original mixing ratio $\rm q_v(p_1)$, since $\rm q_v$ is conserved upon lifting, but the ambient virtual temperature is computed using the local value $\rm q_v(p_2)$, which is generally different from $\rm q_v(p_1)$. For a lifted parcel, neutral stability requires $\rm T_{v,ad}(p_2|p_1) \le T_{v,amb}(p_2)$, or equivalently,
\begin{equation}\label{eq:DisplacedBuoyancy}
    \rm T_{v.amb}(p_1) \left( \frac{p_2}{p_1} \right) ^ {\beta_{mix}(p_1)} \le T_{v,amb}(p_2),
\end{equation}
with the equality defining marginal stability (i.e. neutral buoyancy), in which case the displaced parcel neither rises nor sinks. The criterion for a depressed parcel is the same, except with the inequality reversed.

\subsection{Local stability: The Ledoux criterion}
For stability against infinitesimal displacements, consider $\rm p_2/p_1 = 1-\delta$, with $0<\delta<<1$ for infinitesimal lifting. Substitute $\rm p_2/p_1 = 1-\delta$ in Eqn.~\ref{eq:DisplacedBuoyancy}, Taylor expand, and pass to the limit $\delta \rightarrow 0$ to obtain,
\begin{equation}\label{eq:ledoux_criterion}
    \rm \left.\frac{d\ln T_{v,amb}}{d\ln p}\right\rvert_{p_1} \le \beta_{mix}(p_1).
\end{equation}
The criterion for neutral stability of a slightly depressed parcel ($\rm \delta < 0$) is identical. Eqn.~\ref{eq:ledoux_criterion} is a form of the Ledoux criterion for local stability \citep{Ledoux1947} at $\rm p_1$. The Ledoux criterion has the same form as the Schwartzschild criterion \citep{Schwarzschild1958}, which applies to the compositionally homogeneous case, except with virtual temperature substituted for temperature.

Suppose that the ambient virtual temperature profile is marginally stable (Eqn.~\ref{eq:ledoux_criterion} with inequality replaced by equality) for all $\rm p_1$ within a layer. Then, integrating Eqn.~\ref{eq:ledoux_criterion} over the layer, we find
\begin{equation}\label{eq:LedouxStableProfile}
   \rm T_{v,amb}(p) = T_{v,amb}(p_0) e^{\int_{\ln p_0}^{\ln p}\beta_{mix}(p) d\ln p},
\end{equation}
where $\rm T_{v,amb}(p_0)$ is the virtual temperature at some point $\rm p_0$ within the layer.
Eqn.~\ref{eq:LedouxStableProfile} defines a unique local, marginally stable virtual temperature profile that passes through $\rm (p_0,T_{v,amb}(p_0))$.
However, there is a continuous family of temperature and composition profiles that correspond to the same marginally stable virtual temperature profile, $\rm T_{v,amb}(p)$, given by inserting any end state $\rm q_v(p)$ into Eqn.~\ref{eq:virtual_temp}. This contrasts with the compositionally homogeneous case where there is a unique marginally stable end state passing through a given temperature-pressure point.

The profile defined by Eqn.~\ref{eq:LedouxStableProfile} is not the virtual adiabat that passes through $\rm (p_0,T(p_0))$, since the virtual adiabat would be determined from Eqn.~\ref{eq:VirtualAdiabatExact} with fixed $\rm \beta_{mix}(p_0)$. If $\rm \beta_{mix}(p)$ is independent of $p$, the Ledoux-stable profile reduces to the virtual adiabat. This is the case when the composition is homogeneous in the layer, $\rm q_v(p) = constant$, or for inhomogeneous compositions it can be approximately true if the two gases in the mixture have nearly the same numbers of excited internal degrees of freedom.

\subsection{Stability to finite amplitude displacements}

\begin{figure}
    \centering
    \includegraphics[width=\textwidth]{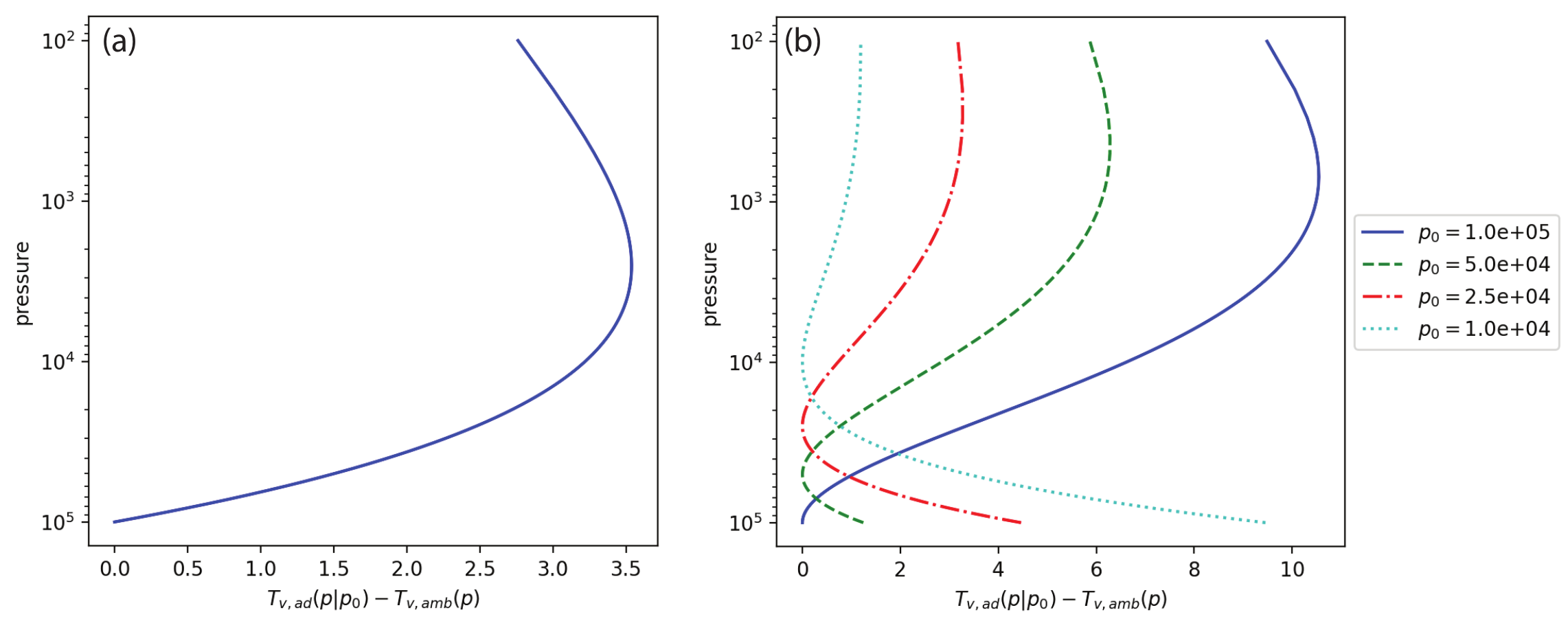}
    \caption{Buoyancy for finite adiabatic displacements from pressure $p_0$ in an ambient profile. The atmosphere is a mixture of $\mathrm{H_2}$ with water vapour, with water vapor profile $q_v(p) = 10 (p/p_0)$, where $p_0$ is the surface pressure. (a) The ambient profile is the virtual adiabat passing through $\rm p= 5\cdot 10^4$ Pa and T = 300 K, and the parcel is lifted from $\rm p_0=10^5$ Pa. (b) The ambient profile is the Ledoux-stable profile $\rm T_{v,ad}(p,\frac{1}{2}p_0)$, with $T_{v,ad} = 300 \; \mathrm{K}$ at $\rm p = \frac{1}{2}p_0$, and the parcel is lifted from the various $p_0$ given in the legend.}
    \label{fig:VirtualAdiabatBuoyancy}
\end{figure}

For compositionally homogeneous atmospheres, if the ambient temperature profile within a given layer is taken to be the dry adiabat passing through any $\rm (p_1,T_1)$ point, then adiabatic displacements of parcels by any amount within the layer yields zero buoyancy, so that the layer is marginally stable to finite amplitude displacements. This is not generally the case for virtual adiabats in a compositionally inhomogeneous atmosphere. Suppose that within a layer, $\rm T_{v,amb}(p) = T_{v,ad}(p|p_1)$ for some $\rm p_1$. If we adiabatically lift a parcel from some pressure $\rm p_0 \ne p_1$ within the layer, without allowing mixing, the buoyancy of the lifted parcel is then proportional to
\begin{equation}
    \rm T_{v,ad}(p|p_0) - T_{v,ad}(p|p_1) = T_{v,ad}(p_0|p_1) \left( \left( \frac{p}{p_0}\right)^{\beta_{mix}(p_0)}-\left( \frac{p}{p_0}\right)^{\beta_{mix}(p_1)} \right).
\end{equation}
Since $\rm \beta_{mix}(p_0) \ne \beta_{mix}(p_1)$ when $q_v(p)$ is non-uniform, the buoyancy of the displaced parcel is generally nonzero, and could be either neutrally stable or unstable depending on the profile of $\rm \beta_{mix}$. Fig. \ref{fig:VirtualAdiabatBuoyancy}(a) illustrates this effect for the case of water vapor mixed with an $\mathrm{H_2}$ background, with concentration profile $\rm q_v(p) = 10 (p/p_0)$, where $p_0$ is the surface pressure. The ambient virtual temperature is taken to be $\rm T_{v,ad}(p,\frac{1}{2}p_0)$, with $T_{v,ad} = 300 \; \mathrm{K}$ at $\rm p = \frac{1}{2}p_0$. The figure shows the buoyancy for a parcel lifted adiabatically without mixing from $p_0$ to $p$. We see that (by construction) a parcel displaced by any amount from $\rm p=\frac{1}{2}p_0$ would have zero buoyancy, but a parcel lifted from $p_0$ is positively buoyant and unstable to convection. Since the vertical variation of $\rm \beta_{mix}$ is small, the buoyancy generated is small, equivalent to under a 3.5 K excess in virtual temperature.

To illustrate that a layer deemed locally marginally stable by the Ledoux criterion is not necessarily stable to finite amplitude displacements in Fig. \ref{fig:VirtualAdiabatBuoyancy}(b), we compute the buoyancy of displacements from various $p_0$ of a Ledoux-stable profile computed from Eqn.~\ref{eq:LedouxStableProfile} with the same composition and $\rm q_v(p)$ as in the preceding example. As expected, the buoyancy vanishes at each $\rm p_0$, but increases only quadratically (as opposed to linearly) with displacement; the latter confirms marginal stability to small displacements. However, for large displacements in the upward or downward directions, buoyancy for this profile is positive: a parcel lifted by a finite amount will continue to rise. A parcel displaced downward by a finite amount will have positive buoyancy and ascend back towards the original position, but thereafter probably overshoot into a positive displacement and continue to rise, thus triggering instability. The instability to finite amplitude displacements arises from $\rm \beta_{mix}(p)$. If the layer were well-mixed, so that $\rm \beta_{mix}$ were constant in the layer, then global marginal stability in the layer would be assured.

\subsection{Virtual potential temperature}

One can define a virtual potential temperature $\theta$ corresponding to adiabatic unmixed displacements from pressure level $p$ to some reference pressure level $p_{00}$:
\begin{equation}\label{eq:PotentialTemperatureExact}
    \rm \theta_v \equiv T_v(p)\left(\frac{p}{p_{00}}\right)^{-\beta_{mix}(p)}.
\end{equation}
Note that in Eqn.~\ref{eq:PotentialTemperatureExact} the exponent $\rm \beta_{mix}$ is evaluated at the parcel's origin $p$, because an unmixed parcel retains the $\rm q_v$, and hence $\rm \beta_{mix}$ of its origin. For a compositionally homogeneous atmosphere, which has uniform $\rm \beta_{mix}$, the choice of reference pressure $\rm p_{00}$ changes the profile of $\rm \theta_v(p)$ by a multiplication constant and does not change the shape of the profile. Uniform $\rm \theta_v$ within a layer indicates global marginal stability to displacements between any pair of points within the layer. However, when $\rm \beta_{mix}$ depends on $p$, a profile of $\rm \theta_v$ that is uniform for one choice of $p_{00}$ will not generally be uniform for a different choice of $p_{00}$. This is another way of phrasing the result illustrated in Fig.~\ref{fig:VirtualAdiabatBuoyancy}. Defining a virtual adiabat based on displacement from one position does not guarantee zero buoyancy when a parcel is displaced from a different position.

We have shown in Fig.~\ref{fig:VirtualAdiabatBuoyancy} that the effects of variation of $\rm \beta_{mix}$ on buoyancy are small. In practice, few inaccuracies are produced by using an approximate virtual temperature based on fixed $\rm \overline{\beta}_{mix}$;
\begin{equation}\label{eq:PotentialTemperatureApproximate}
    \rm \theta_v = T_v(p)\left(\frac{p}{p_{00}}\right)^{-\overline{\beta}_{mix}},
\end{equation}
where $\rm \overline{\beta}_{mix}$ is chosen to be typical of the value of $\rm \beta_{mix}$ within the layer. Constant approximate virtual potential temperature within a layer guarantees approximate marginal stability for finite amplitude adiabatic unmixed displacements between any two points within the layer.

We can also define a virtual potential temperature-like quantity based on the Ledoux-stable profile in Eqn.~\ref{eq:LedouxStableProfile},
\begin{equation}\label{eq:LedouxPotentialtemperature}
    \rm \Tilde{\theta}_v = T_v(p)e^{-\int_{\ln p_{00}}^{\ln p}\beta_{mix}(p) d\ln p}.
\end{equation}
If $\rm \Tilde{\theta}_v$ is constant within a layer, then the layer is marginally stable to infinitesimal displacements starting from any point within the layer. This is the virtual potential temperature referred to in \cite{Leconte2017}. However, Eqn.~\ref{eq:LedouxPotentialtemperature} is not a true potential temperature in the usual sense, because it does not correspond to the correct temperature change upon adiabatic unmixed displacements. Uniformity of $\rm \Tilde{\theta}_v$ does not guarantee global marginal stability. \cite{Leconte2017} do not discuss this limitation. In practice, the distinction between Eqn.~\ref{eq:LedouxPotentialtemperature} and a true virtual potential temperature is largely immaterial, because Eqn.~\ref{eq:LedouxPotentialtemperature} reduces to Eqn.~\ref{eq:PotentialTemperatureApproximate} when $\rm \beta_{mix}$ is constant, or the variations in $\rm \beta_{mix}$ are generally too small to yield important effects.

\subsection{CAPE analysis} \label{sec:atmosstability}

We use convective available potential energy (CAPE) to assess the stability of an atmosphere to finite amplitude displacements and to determine the amount of energy available for mixing. CAPE represents the work per unit mass the buoyancy force exerts to lift an air parcel containing a mixture of an atmospheric tracer and background air \citep[][Chap.6]{Emanuel1994}. CAPE is defined as,
\begin{equation} \label{eq:cape}
    \rm CAPE = \int_{LNB}^{p_{init}} R_{b}\left[ T_{v,parcel}(p) - T_{v, amb}(p)\right] dlnp,
\end{equation}
where $\rm p_{init}$ is the initial parcel location, LNB is the level of neutral buoyancy, $\rm T_{v,parcel}(p)$ is the virtual temperature profile of a rising air parcel containing an atmospheric tracer, and $\rm T_{v, amb}(p)$ is the virtual temperature profile of the ambient atmosphere. Eqn.~\ref{eq:cape} assumes there is no condensation or retained condensates. The LNB is the pressure at which a rising air parcel has zero buoyancy and the $\rm T_{v,parcel}$ is equal to $\rm T_{v, amb}$. When a rising air parcel has positive CAPE it will freely rise within the atmosphere beyond the LNB to the Level of Maximum Ascent (LMA) which is when the kinetic energy of the rising air parcel becomes zero. After the rising air parcel reaches the LMA, the parcel would oscillate around the LNB until, under the action of friction, it settles at some point near the LNB (i.e. neutral stability). Note that the temperature of the lifted parcel is given by the virtual adiabat in Eqn. \ref{eq:VirtualAdiabatExact}, which is computed assuming the parcel retains the value of $\rm \beta_{mix}$ of its origin.

Analysing the CAPE of air parcels lifted by finite amounts from various heights within the atmospheric allows one to determine the global stability of the atmosphere. Regions of positive CAPE are convectively unstable and will mix if perturbed. Regions of negative CAPE are stably stratified. If an air parcel with negative CAPE is perturbed upwards, it will sink within the atmosphere to its initial location unless it has enough external energy to rise above the region of negative CAPE. Regions of negative CAPE usually imply that convection will be inhibited unless there is an external energy source \citep[][Chap.6]{Emanuel1994}. Finally, when CAPE = 0, the system is marginally stable.

In $\rm H_2$ atmospheres, atmospheric tracers have a stabilizing effect on convection and can produce a stable layer within these planetary atmospheres \citep{Li2015, Leconte2017}. \cite{Li2015} showed that when a compositionally stable layer is heated to overcome the convective stability due to an atmospheric tracer, the atmosphere can experience giant storms. \cite{Leconte2017} explored the effect a compositionally stable layer would have on the atmospheric state using 1D analytical models. They argued that the only mechanism to transport heat into a stably stratified compositional layer is by radiation, which leads to hotter temperatures in the lower atmosphere, and thus impacts our understanding of planetary atmospheric chemistry and dynamics.

We aim to explore how compositional gradients affect the end state of convective mixing. For a compositionally homogeneous atmosphere, convection mixes the final atmospheric state to a dry adiabat which is uniquely defined once the boundaries of the mixed layer are determined. The final atmospheric state after convective mixing in a compositionally homogeneous situation is a global, marginally stable state.

For a compositionally inhomogeneous atmosphere, if the final mixed state has remaining compositional inhomogeneities, the most general marginal stability condition that can be demanded is that the end state profiles satisfy the Ledoux local marginal stability condition everywhere. The corresponding virtual temperature profile is given by Eqn.~\ref{eq:LedouxStableProfile}. However, as shown in the discussion above, the effect of the variation of $\beta_{mix}$ on buoyancy is small even when there are considerable variations in $\rm q_v$. Therefore, the virtual adiabats are nearly the same regardless of which level a parcel is displaced from. Similarly, the marginally Ledoux stable profile is nearly marginally stable to finite amplitude displacements. We have found that taking into account variations in $\rm \beta_{mix}$ yields only small changes (invisible for graphical purposes) in the stabilized virtual temperature profile as compared to fixing $\rm \beta_{mix}$ at some value representative of its typical value within the mixed region. Thus, we define an approximate virtual adiabat with fixed $\rm \beta_{mix}$, even if the mixed state has some remaining compositional variations. Similarly, we can define an approximate virtual potential temperature whose constancy within the mixed region defines global marginal stability. In the subsequent discussion, we will drop the term ``approximate" and simply refer to ``virtual adiabat" and ``virtual potential temperature." Note that the fixed $\rm \beta_{mix}$ approximation does not eliminate the degeneracy in the family of final marginally stable temperature and composition profiles. Only the marginally stable virtual temperature profile is uniquely determined by a virtual adiabat. The effect of $\rm q_v$ on temperature for fixed virtual temperature is still large.

We use 3D convection resolving simulations to address the following questions within this work:
\begin{enumerate}
 \item What is the final compositional profile after mixing in an initially inhomogeneous composition atmosphere?

 \item Does convection mix the final atmospheric state in an initially inhomogeneous composition atmosphere to a local or global marginally stable state?

 \item If convection mixes the atmosphere to a locally stable state, can it be described by the virtual adaibat?

 \item How do large-scale versus small-scale initial perturbations affect the final state of convective mixing?

\end{enumerate}
Knowing how convection mixes the compositional profile is important to discern which among the many possible final state profiles is realized.

\section{Compositional Convection Modelling with Cloud Model 1} \label{sec:model_description}

\begin{table*}[ht!]
	\caption{Atmospheric constants used within the CM1 simulations}
	\label{table:atmospheric_composition}
	\centering
	\begin{tabular}{lllll}
		\toprule
		{} & {Cp} & {Cv} & {R} & {$\rm \varepsilon = M_{v}/M_b$} \\
		{} & {J/(kg K)} & {J/(kg K)} & {J/(kg K)} & {} \\
		\midrule
		Earth-air & 1005.7 & 718.6 & 287.0 & 0.621 \\
		Hydrogen ($\rm H_2$) & 14304.0 & 10160.0 & 4124.2 & 8.94 \\
		Carbon Dioxide ($\rm CO_2$) & 844.0 & 655.0 & 188.9 & 0.409 \\
		Water Vapour (Atmospheric tracer) & 1870.0 & 1408.5 & 461.5 & N/A \\
		\bottomrule
	\end{tabular}
\end{table*}

\begin{table*}[th!]
	\caption{Initial atmospheric sounding parameters and domain size used for the CM1 initial value problem simulations. Cases 1 - 6 list the CM1 parameters for the high resolution simulations. Cases 7 - 12 and list the CM1 setup for the parameter study simulations. Cases 1 - 3 use sounding profile 1, and Cases 4 - 12 use sounding profile 2. $\rm z_1$ and $\rm z_2$ represent the heights specified within CM1 at which the density discontinuity starts and ends, corresponding to $\rm p_1$ and $\rm p_2$ respectively. The domain is provided in km, dx and dy are the horizontal and vertical resolution respectively.}
	\label{table:initial_values}
	\centering
	\begin{tabular}{lcccccccccll}
		\toprule
		{CM1 Testcase} 	& {$\rm T_s$} 	& {$\rm T_1$} 	& {$\rm q_{v,s}$} 	& {$\rm q_{v,1}$} 	& {$\rm z_1$} 	& {$\rm z_2$} 	& {Domain} 	& {dx} & {dz} \\
		{} & {K} & {K} & {kg/kg} & {kg/kg} & {m} & {m} & {(Lx, Ly, Lz)} & {m} & {m} \\
		\midrule
 1 Isothermal Earth Air & 450 & 450 & 0.5 & 0.0 & 18638 & 24863 & (30, 30, 75) & 150 & 75 \\
 2 Isothermal $\rm H_2$ & 450 & 450 & 0.0 & 0.5 & 216195 & 288405 & (400, 400, 870) & 2000 & 870 \\
 3 Isothermal $\rm CO_2$ & 450 & 450 & 0.5 & 0.0 & 18638 & 24863 & (30, 30, 75) & 150 & 75 \\
 4 Non-isothermal Earth Air & 450 & 400 & 0.5 & 0.0 & 18638 & 24863 & (30, 30, 75) & 150 & 75 \\
 5 Non-isothermal $\rm H_2$ & 700 & 400 & 0.5 & 0.0 & 216195 & 288405 & (400, 400, 870) & 2000 & 870 \\
 6 Non-isothermal $\rm CO_2$ & 450 & 400 & 0.3 & 0.0 & 18638 & 24863 & (30, 30, 75) & 150 & 75 \\
 7 CTE $\rm T(p)$ Earth Air & 450 & 400 & 0.1 - 0.7 & 0.0 & 18375 & 24625 & (300, 300, 75) & 1500 & 250 \\
 8 CTE $\rm T(p)$ $\rm H_2$ & 700 & 400 & 0.1 - 0.7 & 0.0 & 218250 & 290250 & (1200, 1200, 900) & 6000 & 4500 \\
 9 CTE $\rm T(p)$ $\rm CO_2$ & 450 & 400 & 0.1 - 0.5 & 0.0 & 18375 & 24625 & (300, 300, 75) & 1500 & 250 \\
 10 CTE $\rm q_v(p)$ Earth Air & 400 & 450, 500, 550 & 0.3 & 0.0 & 18375 & 24625 & (300, 300, 75) & 1500 & 250 \\
 11 CTE $\rm q_v(p)$ $\rm H_2$ & 700 & 500, 600, 700 & 0.3 & 0.0 & 218250 & 290250 & (1200, 1200, 900) & 6000 & 4500 \\
 12 CTE $\rm q_v(p)$ $\rm CO_2$ & 450 & 450, 500, 550 & 0.3 & 0.0 & 18375 & 24625 & (300, 300, 75) & 1500 & 250 \\
		\bottomrule
	\end{tabular}
\end{table*}

Cloud Model 1 (CM1; \citealt{Bryan2009}) is a numerical model in 3D Cartesian coordinates that is designed to study microscale (1 – 1000 m) and mesoscale (1 – 1000 km) convective phenomena in idealized simulations. The full equations used within CM1 are described in \citep{Bannon2002, Bryan2009, Bryan2012}.

CM1 is originally developed for Earth convection studies. To accommodate modelling exoplanet atmospheres, we made two modifications to the CM1 code. First, we modified CM1 to allow the user to change the composition of both the bulk atmosphere and the atmospheric tracers. Numerically, the atmospheric composition defines the thermodynamic constants such as the ideal gas constant ($\rm R$), specific heat capacity at constant pressure ($\rm C_p$) and constant volume ($\rm C_{p,v}$), as well as, any latent heating terms for the atmospheric tracer which are used in the equations of motion. To allow the user to specify thermodynamic gas constants, we modified CM1 to read a user generated input file which lists the planetary atmospheric parameters that are dependent upon composition for both the background atmosphere and atmospheric tracer. Within this work, we consider Earth-air, $\rm H_2$, and $\rm CO_2$ atmospheres with a $\rm H_2O$ vapour non-condensing atmospheric tracer. The gas specific parameters that are used within this study for the three atmospheric compositions are listed in Table \ref{table:atmospheric_composition}.

The second modification made to the CM1 code was to accommodate a non-dilute mass mixing ratio. In CM1, the buoyancy equation assumes a dilute mass mixing ratio \citep[see][]{Bryan2009}. To accommodate non-dilute values, we modified the CM1 code to use a more general form of the buoyancy (B) given by, $\rm B = - g (\rho'_b/\rho_{o, b})$, where B is the buoyancy, $\rm \rho'_{b}$ is the deviation of the background gas density from a reference hydrostatic state given by $\rm \rho_{o, b}$, and g is the planetary gravity \citep{Emanuel1994}.

We performed simplified initial value problem simulations with CM1. All of our CM1 test cases are performed without any condensation/latent heat release, cloud microphysics, or radiation. We can better understand the underlying fluid dynamics of compositional effects on convection by first exploring compositional convection in its simplest form. Non-condensing compositional convection is relevant to planetary atmospheres that are hot enough to suppress condensation. The CM1 simulations presented here create a foundation for future work to explore the behaviour of compositional convection while accounting for condensation and radiation. We plan to perform additional simulations to further characterize the behaviour of compositional convection while accounting for these processes, but focus on presenting the results of our non-condensing convection simulations within this work.

For simplicity, the planetary radius and gravity used within the CM1 simulations are the given Earth values. While these parameters are based on Earth, they do not impact the final results of our simulations. The domain of our convection box is small enough that spherical geometry does not have a significant effect. The surface pressure is set to $10^5$ Pa within all of the CM1 test cases, and there are no initial winds. To explore the effect of compositional convection, we test the behaviour of a non-condensing water vapour atmospheric tracer in three different planetary atmospheres of Earth-air, $\rm H_2$, and $\rm CO_2$ composition with two different initial sounding states described by:
\begin{enumerate}
	\item An isothermal atmosphere at $\rm~T=~T_s$ with a step mass mixing ratio profile of a non-condensing $\rm H_2O$ tracer given by,
	\begin{equation*}
		\rm q_v(p)=\begin{cases}
			\rm q_{v,s} &\text{when} \; \rm{p_s~>~p~>~p_1} \\
			\rm 0.0 	 &\text{when} \; \rm{p_2~>~p~>~p_{TOA}}.
		\end{cases}
	\end{equation*}
	\item An initial atmospheric sounding state with both a temperature and mass mixing ratio step profile given by,
	\begin{equation*}
		\rm q_v(p) =\begin{cases}
			\rm q_{v,s} &\text{when} \; \rm{p_s~>~p~>~p_1} \\
			\rm 0.0 	 &\text{when} \; \rm{p_2~>~p~>~p_{TOA}} \\
		\end{cases}
	\end{equation*}
	\begin{equation*}
		\rm T(p)=\begin{cases}
			\rm T_s &\text{when} \, \rm{p_s~>~p~>~p_1} \\
			\rm T_1 	 &\text{when} \, \rm{p_2~>~p~>~p_{TOA}}.
		\end{cases}
	\end{equation*}
\end{enumerate}

$\rm p_1$ and $\rm p_2$ are arbitrarily chosen pressure levels taken in the lower part of the atmosphere with $\rm p_2~<~p_1$. The discontinuity locations $\rm p_1$ and $\rm p_2$ are specified in the CM1 code to occur at p(nk/4) and p(nk/3) respectively, where nk is the number of vertical grid points, which is given by Lz / dz. In both initial sounding profiles, a high density atmospheric layer is lying above a lower density atmospheric layer. Both initial sounding states are Rayleigh-Taylor unstable. In sounding profile 1, only the compositional gradient is creating a density difference between the upper and lower atmospheric layers. While in sounding profile 2, the temperature and compositional profiles contribute to the density difference between the upper and lower atmosphere. Between $\rm p_1$ and $\rm p_2$ the mass mixing ratio and temperature (when applicable) profiles linearly vary with height. A linear gradient is used to smear out the density profile (i.e. prevent a sharp discontinuity in density at the step interface) to make the initial atmospheric state more stable and allow the CM1 code to run to convergence without becoming numerically unstable. For a step interface in a Rayleigh-Taylor unstable system, the growth rate of the instability increases indefinitely as the wavelength of the perturbation decreases. In numerical simulations this leads to unresolved grid-scale motions if not suppressed. By using a smeared out density transition within the CM1 simulations, a shortwave cutoff is introduced \citep{bud1992stabilization}, which prevents unresolved scales from emerging. In all of the simulations performed within this work, the Mach number is small. Therefore, while compressibility affects the temperature, sound wave and shock generation does not occur to any appreciable extent.

We perform two sets of CM1 simulations. First, we run a series of high resolution CM1 simulations, and second, we perform a parameter study in which CM1 is run with a lower resolution compared to the high resolution simulations. Even though we do not resolve the smallest scale plumes in the parameter study, the final atmospheric profiles of the parameter study are consistent with the results from the high resolution simulations. For all of the CM1 test cases performed, we use doubly periodic horizontal boundary conditions, and large eddy simulation mode with the turbulent kinetic energy (TKE) sub-grid turbulence scheme \citep{Deardorff1980}. We use the TKE scheme to represent turbulent diffusion on a sub-grid scale and allow for mixing within the CM1 simulations. Turbulent mixing will increase the entropy of the system relative to the initial atmospheric state. The turbulence parameterization represents the cascade of composition and potential temperature variability down to scales where molecular diffusion provides the final stage of mixing that ultimately leads the entropy to increase.

Additionally, we implemented a Rayleigh dampening layer in the top 7000 m of the vertical domain with a dampening coefficient of $\rm \alpha=1/300$ s to damp out upward propagating gravity waves. Table~\ref{table:initial_values} lists the initial parameters for all CM1 test cases explored in this study. For simplicity, we label each test case with a number which is listed in Table~\ref{table:initial_values}.

\subsection{High Resolution CM1 Simulations}
For the high-resolution simulations, we use $\rm q_{v,s} \, = $ 0.5 kg/kg, $\rm T_s = 450$ K and $\rm T_1$ = 400 K for the Earth-air atmosphere case with both sounding profiles and the $\rm CO_2$ atmospheres with sounding profile 1. For the $\rm CO_2$ atmosphere case with sounding profile 2, having $\rm q_{v,s} \, = $ 0.5 kg/kg caused the full atmospheric domain to mix which led CM1 to become numerically unstable in the high resolution simulations. With $\rm q_{v,s} \, = $ 0.5 kg/kg and sounding profile 2, we observed numerical artifacts from the interaction of convection with the upper boundary of the simulation domain. For this case, we set $\rm q_{v,s} \, = $ 0.3 kg/kg to run CM1 in a more stable configuration with $\rm T_s = 450$ K and $\rm T_1$ = 400 K.

For the isothermal hydrogen atmosphere test case (sounding profile 1), having a water vapour tracer in the lower part of a hydrogen atmosphere is convectively stable. Therefore, we initially place the water vapour tracer in the upper atmosphere with no $\rm H_2O (v)$ in the lower part of the atmosphere and use a step mass mixing ratio profile given by $\rm q_{v}~=~0$ from $\rm p_s~>~p~>~p_1$ and $\rm~q_v~=0.5$ from $\rm p_2~>~p~>~p_{TOA}$. Again, a linear relationship between mass mixing ratio and height is used to smear out the density interface between $\rm p_1$ and $\rm p_2$. Secondly, for the non-isothermal hydrogen atmosphere test case (sounding profile 2), we use $\rm T_s = 700~K$ and $\rm T_1$ = 400 K to create positive buoyancy of the atmospheric parcels in the lower half of the atmosphere where $\rm H_2O$ is present with $\rm q_v\,=\,0.5$. Note, here the absolute value of temperature does not affect the simulation results since the temperature and mass mixing ratio are set independently of each other. We chose to set the minimum atmospheric temperature at 400 K since water does not condense at this temperature. For the $\rm H_2$ atmosphere the lower atmosphere temperature was set to be the first temperature at which we observed any significant convection within the CM1 simulation results when $\mathrm{q_{v,s} = 0.5}$.

The CM1 domain for the Earth-air and $\rm CO_2$ atmosphere is (Lx, Ly, Lz) = (30, 30, 75) km, and for the $\rm H_2$ atmosphere is (Lx, Ly, Lz) = (400, 400, 870) km. The vertical domain height was set such that $\rm p_{TOA}$ is approximately $\mathcal{O}(10^3)$ Pa and the vertical domain is representative of an atmospheric column. A vertical resolution of 75 m and horizontal resolution of 150 m is used for the Earth-air and $\rm CO_2$ atmosphere simulations. For the $\rm H_2$ atmosphere, we used a 870 m and 1000 m vertical and horizontal resolution respectively.

Finally, for all six high resolution CM1 test cases, we explored the effect of two different initializations. First, we ran all of the test cases with initial small-scale random perturbations in potential temperature ($\theta$) with an amplitude of $\pm$~0.5~K. Second, we performed the CM1 test cases initialized with both random small-scale perturbations and a large-scale sinusoidal perturbation of the form,
\begin{equation} \label{eq:3d_wave}
    \rm \theta' = sin\left(\frac{2 \pi x}{0.75 W}\right) sin\left(\frac{2 \pi y}{0.75 W}\right) sin\left(\frac{2 \pi z}{2H}\right),
\end{equation}
where $\rm \theta'$ is the potential temperature perturbation, W is the width CM1 horizontal domain for the given test case, and H is a vertical scale height which is set to 50 km for the Earth-air and $\rm CO_2$ atmosphere and 500 km for the $\rm H_2$ atmosphere test cases. Using only random perturbations in $\theta$ produces small-scale vertical displacements of the initial state, which allows us to explore the Ledoux criterion condition with CM1 (Eqn.~\ref{eq:ledoux_criterion}). When using random small-scale initial perturbation, the shortest wavelengths resolved are the most unstable and will lead to the formation of small-scale turbulent plumes. Conversely, the sinusoidal perturbation in $\theta$ creates a large, finite displacement of an atmospheric air parcel. Random perturbations were added in conjunction with the sinusoidal wave to add variation to the horizontal domain. When the initial condition is dominated by a large-scale perturbation, a large-scale coherent plume has the potential to grow to a finite amplitude before the small-scale instabilities take over. Using a large-scale perturbation allows us to explore compositional convection with finite amplitude vertical displacements of an air parcel, test the sensitivity of the final state to spatial scales excited in the system, and evaluate whether the Ledoux criterion is still sufficient in describing the final atmospheric state.

All CM1 cases were run long enough for the atmospheric state to converge to a final marginally stable state. The Earth-air and $\rm CO_2$ atmosphere were run for 3600 seconds, while the $\rm H_2$ atmosphere test cases were run for 8640 seconds.

For CM1 Case 1--3, which use sounding profile 1, the compositional gradient creates atmospheric instability which overcomes the stable isothermal temperature profile. For test case 4 and 6, both the temperature and compositional gradient render the initial profile unstable to convection. For Case 5, the $\rm H_2$ atmosphere with sounding profile 2, the compositional gradient stabilizes the atmosphere while the temperature gradient creates an initial profile that is unstable to convection. Case 1, which models an isothermal Earth-air atmosphere, is similar to a ``hot Earth" but with a greater compositional enhancement of buoyancy due to the larger value of $\rm q_v$ considered in this work. This case is relevant to planets with $\rm N_2$ background atmospheres over an ocean, which are much hotter than Earth and therefore have a higher $\rm q_v$. Hot Earth planets would be saturated and thus have condensation and latent heat release. Our non-condensing calculation does capture part of the hot Earth case, but latent heat release would make convective mixing more vigorous. Case 3, the isothermal $\rm CO_2$ atmosphere, would be relevant for terrestrial exoplanets with a higher molecular weight atmosphere over a $\rm CO_2$ ocean. Lastly, the hydrogen atmosphere cases look to explore compositional effects for sub-Neptune exoplanet atmospheres. Even though the simulations here do not fully represent the exoplanets we aim to study, they are important because they illustrate the role compositional buoyancy would play in such atmospheres.

\subsection{CM1 Parameter Study}
The parameter study explores how varying the initial compositional and temperature gradient respectively would affect the final atmospheric profiles after mixing. The parameter study is performed using sounding profile 2. In the parameter study, we run a set of simulations where we (1) varied the initial lower atmospheric mass mixing ratio value, $\rm q_{v, s, initial}$, while keeping the temperature profile constant; (2) varied the initial lower atmosphere $\rm T_s$ value while keeping the $\rm q_{v}$ profile fixed. In CM1, the start of the smeared out discontinuity location is fixed to occur at a given height. Since both composition and temperature contribute to density, the start of the density discontinuity occurs at different pressure levels for the same given height within a set of simulations for a given atmosphere.

For the constant T(p) parameter study simulations, we vary $\rm q_{v, s, initial}$ from 0.1 to 0.7 in increments of 0.2. For the $\rm CO_2$ atmosphere, the test case with $\rm q_{v, s, initial}$ = 0.7 became numerically unstable and is excluded from the results. For the Earth-air and $\rm CO_2$ atmospheres, T(p) is given by a step profile described by sounding profile 2 with $\rm T_{s, initial}$ = 450 K and $\rm T_{1, initial}$ = 400 K. For the $\rm H_2$ atmosphere, $\rm T_{s, initial}$ = 700 K and $\rm T_{1, initial}$ = 400 K.

For the constant $\rm q_v$(p) parameter study simulations, we vary $\rm T_{s, initial}$ from 400 to 550 K in increments of 50 K for the Earth-air and $\rm CO_2$ atmospheres simulations, and vary $\rm T_{s, initial}$ from 500 to 700 K in increments of 100 K for the $\rm H_2$ atmosphere simulations. For all three atmospheric compositions, $\rm T_{1, initial}$ = 400 K and the mass mixing ratio profile is given by the step profile described by sounding profile 2 where $\rm q_{v, s, initial}$ = 0.3 and $\rm q_{1, initial}$ = 0.0.

The CM1 simulation setup for the parameter study is summarized by Cases 7 - 12 in Table~\ref{table:initial_values}. We employ a CM1 domain for the Earth-air and $\rm CO_2$ atmospheres with dimensions (Lx, Ly, Lz) = (300, 300, 75) km with a horizontal resolution of 1500 and vertical resolution of dz = 250 m. For the $\rm H_2$ atmosphere, the domain is (Lx, Ly, Lz) = (1200, 1200, 900) km with a horizontal resolution of 6000 m and vertical resolution of 4500 m. All CM1 simulations in the parameter study were run with initial small-scale random perturbations to trigger convection, and the simulations were run for 8640 seconds. Due to the lower resolution employed in the parameter study, we do not fully resolve the smallest scale plumes that form at the initial stages of the simulation. Again, we do find that even with the lower resolution the final atmospheric state matches the corresponding high resolution simulations. We reduced the resolution of these runs to save computational time.

\section{CM1 Simulation Results \& Discussion} \label{sec:cm1_results}

\begin{figure*}[ht]
	\centering
	\includegraphics[width=\textwidth]{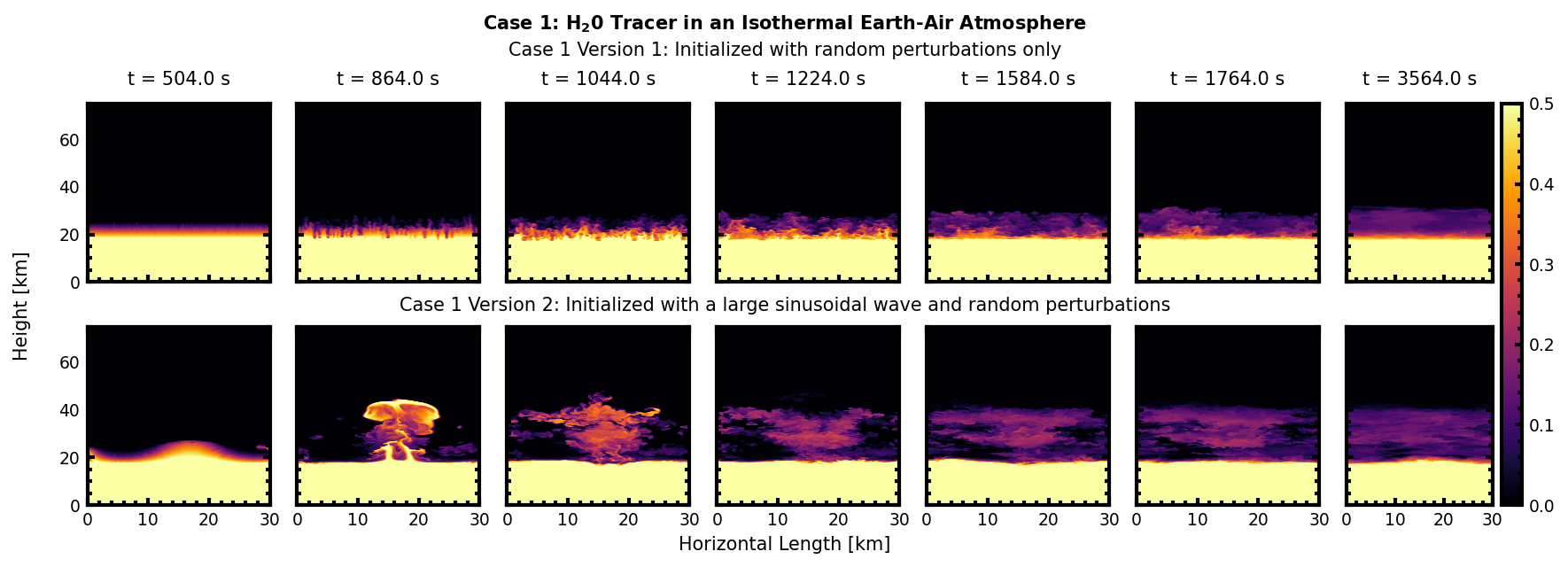}
	\includegraphics[width=\textwidth]{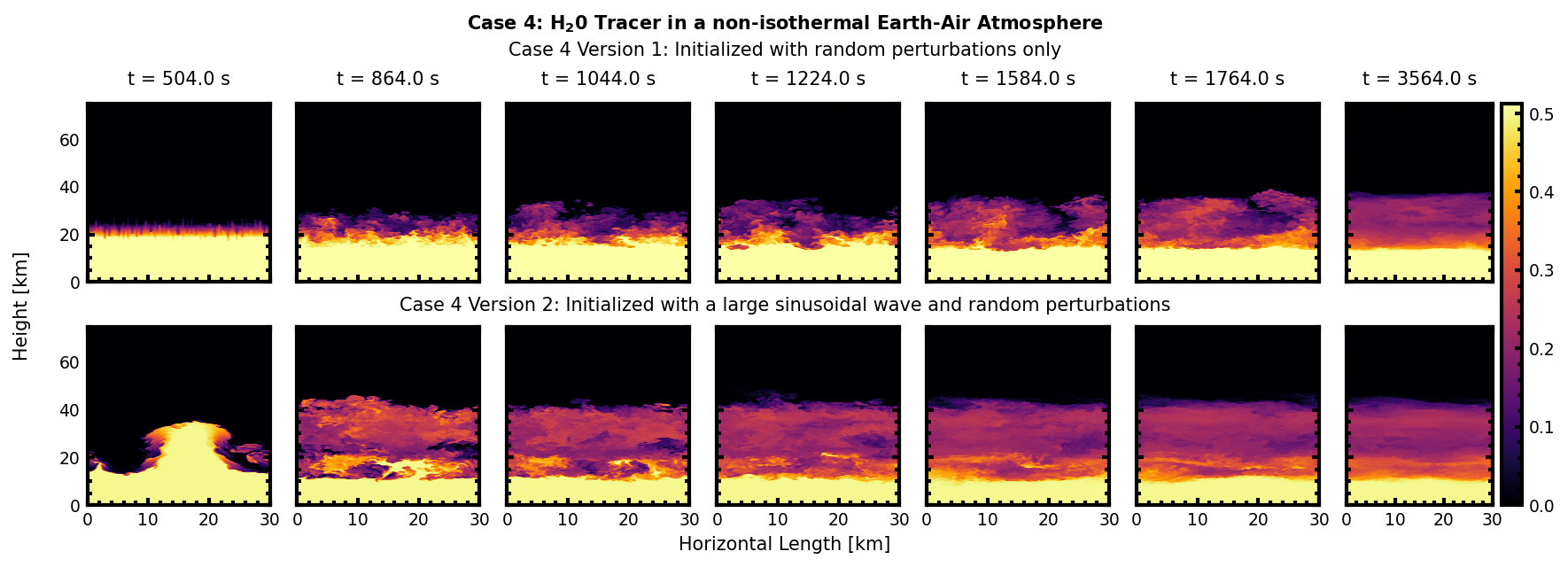}

	\caption{2D plots showing vertical cross sections of mass mixing ratio taken in the middle of the y-domain for CM1 Case 1 and Case 4 which have an Earth Air Atmosphere. The top panel shows the $\rm q_v$ cross sections for Case 1, isothermal Earth air atmosphere, while the bottom panel shows the $\rm q_v$ cross section for Case 4, Earth-Air atmosphere with initial temperature and mass mixing ratio step profile. Within each panel, the top row shows the results when using only initial random perturbations to trigger convection, while the bottom row shows the results when using the initial perturbation given by Eqn.~\ref{eq:3d_wave} in conjunction with random perturbations. The mass mixing ratio is shown for seven different time steps during the CM1 simulation. The color bar indicates the value of $\rm q_v$. The plots show the formation, growth, and mixing of convective plumes. Eventually, the atmosphere reaches a marginally stable state where discrete layers of varying composition form. The observed ``compositional staircases" are analogous to the compositional staircases presented in compositional convection modelling for stellar atmospheres \citep{Garaud2015}.}
	\label{fig:air_2d_mass_mixing_ratio}
\end{figure*}

\begin{figure*}[ht]
	\centering
	\includegraphics[width=\textwidth]{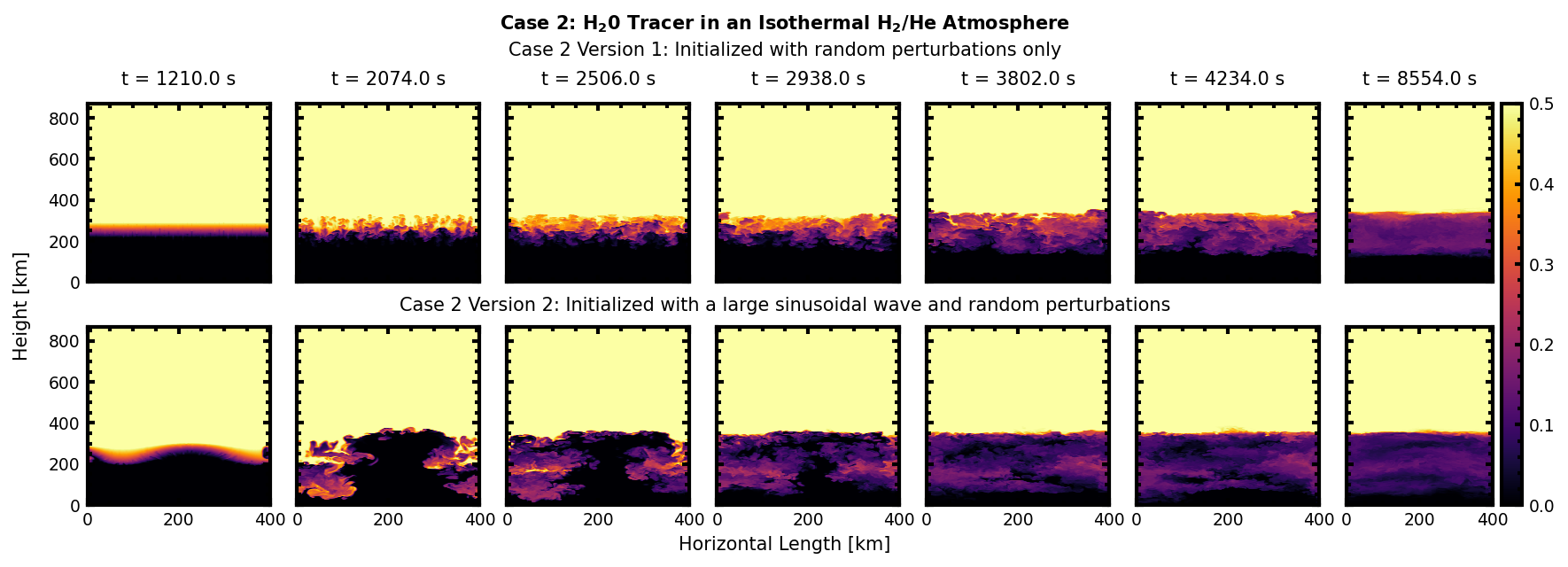}
	\includegraphics[width=\textwidth]{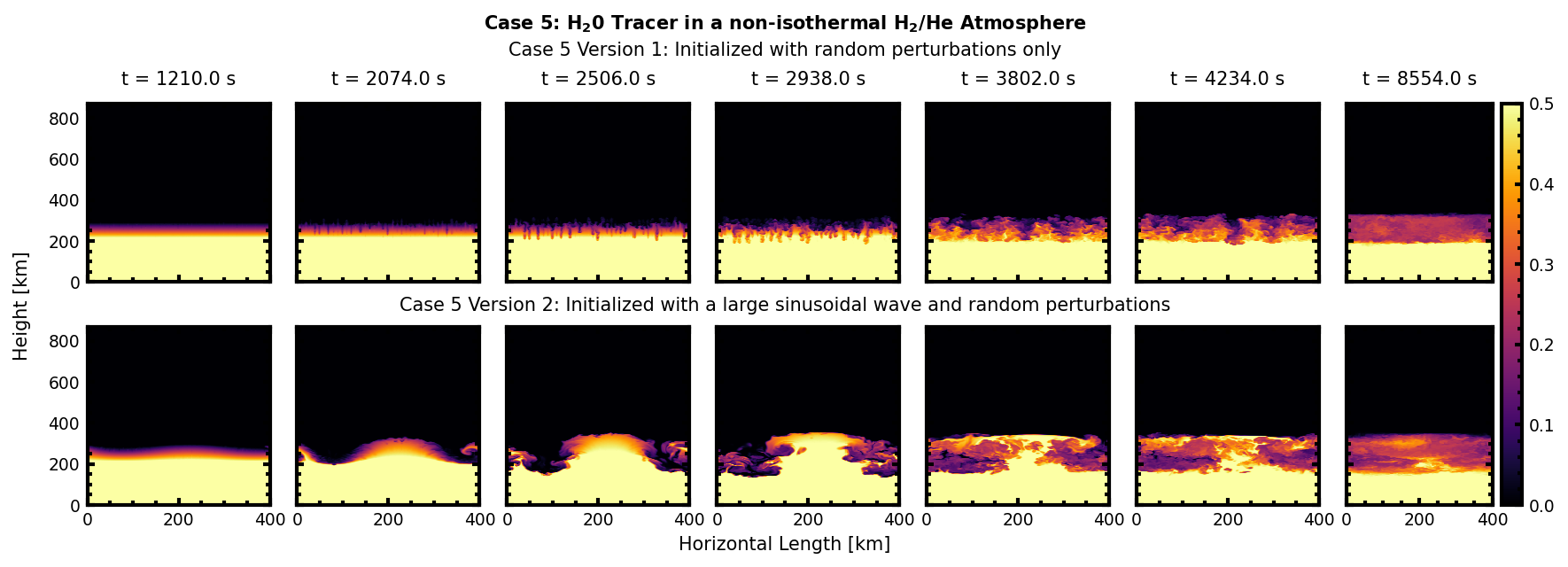}

	\caption{Same as Fig.~\ref{fig:air_2d_mass_mixing_ratio} but for the $\rm H_2$ atmosphere test cases. The top panel shows 2D vertical cross sections of the mass mixing ratio taken in the middle of the y-domain for Case 2, while the bottom panel shows the results for Case 5.}
	\label{fig:h2_2d_mass_mixing_ratio}
\end{figure*}

\begin{figure*}[ht]
	\centering
	\includegraphics[width=\textwidth]{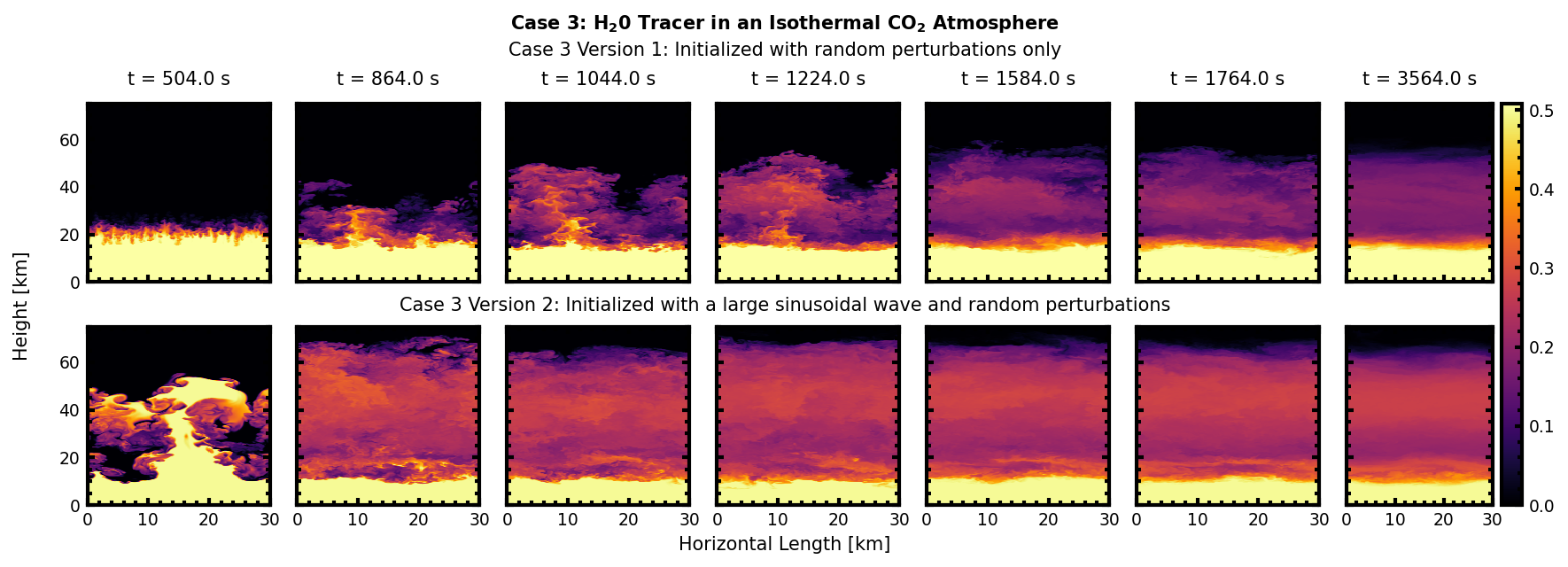}
	\includegraphics[width=\textwidth]{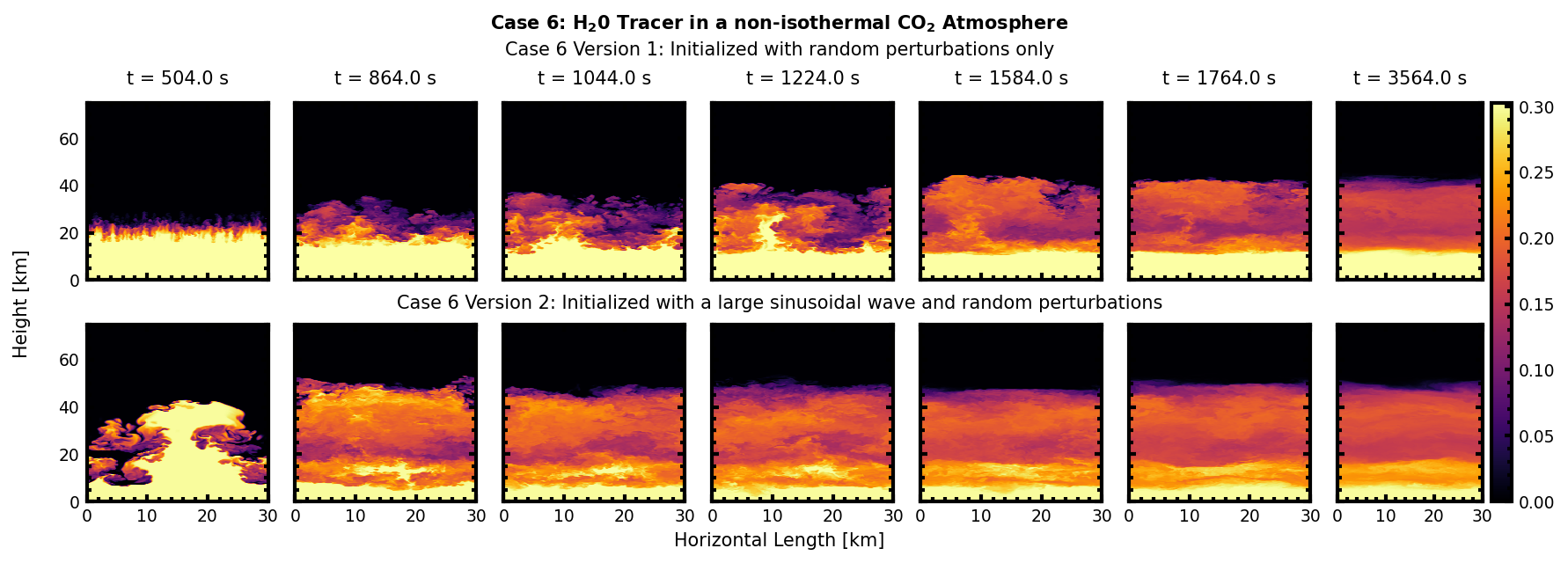}

	\caption{Same as Fig.~\ref{fig:air_2d_mass_mixing_ratio} but for the $\rm CO_2$ atmosphere test cases. The top panel shows 2D vertical cross sections of the mass mixing ratio taken in the middle of the y-domain for Case 3, while the bottom panel shows the results for Case 6. Note that in Case 6, $\rm q_v,s = 0.3$ kg/kg, and thus the color bar for Case 6 is not equivalent to the color bar shown for Case 3.}
	\label{fig:co2_2d_mass_mixing_ratio}
\end{figure*}

\begin{figure*}[ht!]
	\centering
	\includegraphics[width=\textwidth]{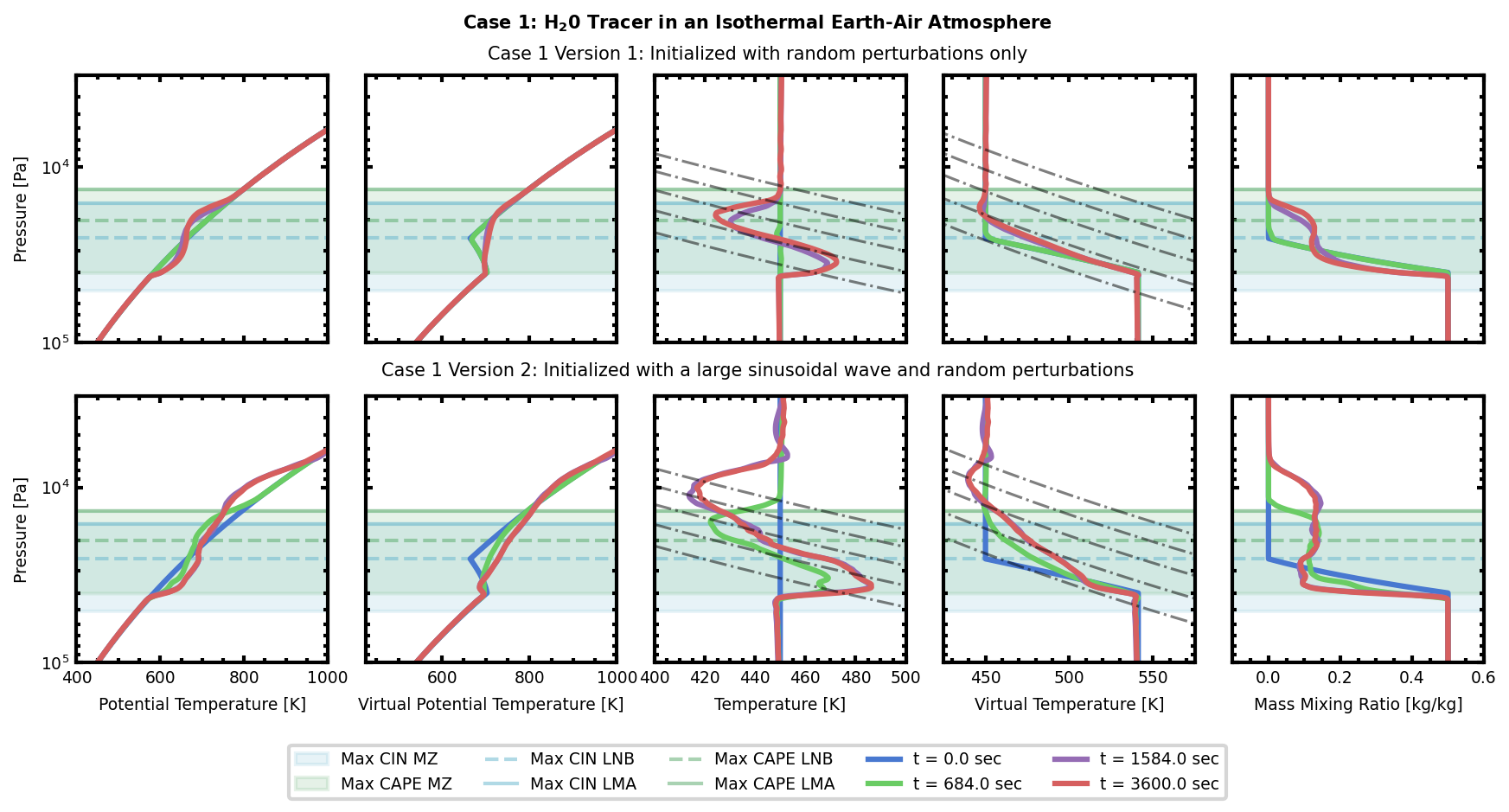}

	\caption{Vertical profiles from left to right of $\rm \theta (p)$, $\rm \theta_v(p)$, $\rm T(p)$, $\rm T_v(p)$, and $\rm q_v(p)$ with respect to pressure for Case 1, isothermal Earth-air atmosphere with an initial non-condensing $\rm H_2O$ in the lower half of the atmosphere. The vertical profiles for $\rm \theta (p)$, $\rm T(p)$, and $\rm q_v(p)$ are determined by taking the horizontal average of the output 3D CM1 data. $\rm T_v(p)$ and $\rm \theta_v(p)$ are calculated using horizontally averaged $\rm T(p)$ and $\rm q_v(p)$ profiles. The coloured lines represent the atmospheric vertical profiles at different four time steps during the CM1 simulation where the initial state is given by a blue line and the final state is the red line. The blacked dashed lines show uniform composition dry adiabats, calculated using Eqn.~\ref{eq:DryAdiabat},in the temperature panel, and virtual adiabats, calculated using Eqn.~\ref{eq:VirtualAdiabatExact}, in the virtual temperature panel. The blue-green shaded region shows the predicted mixing zone from a CAPE analysis performed on the initial atmospheric sounding state using Eqn.~\ref{eq:cape}. Finally, the top row shows the vertical profiles when CM1 Case 1 was run using only random perturbations to trigger convection, while the bottom row shows the results when convection is triggered with both random perturbations and a large sinusoidal wave given by Eqn.~\ref{eq:3d_wave}. Case 1 version 2 shows the least agreement of where convection occurred to the predicted mixing zone.}
	\label{fig:case1_verticalprofiles}
\end{figure*}

\begin{figure*}[ht]
	\centering
	\includegraphics[width=\textwidth]{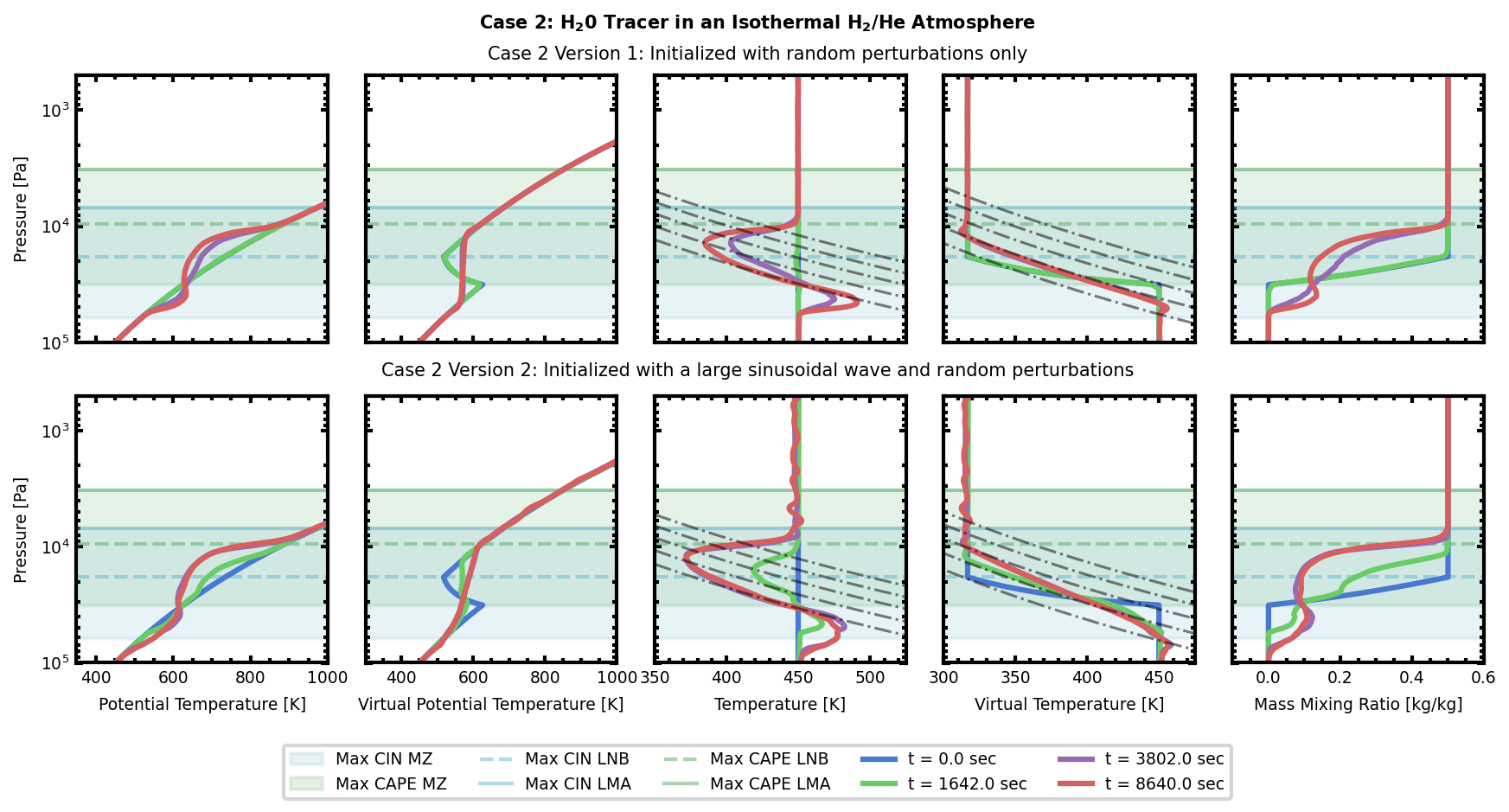}

	\caption{Same as Fig.~\ref{fig:case1_verticalprofiles} but showing the CM1 vertical profiles for for Case 2, isothermal $\rm H_2$ atmosphere with an initial non-condensing $\rm H_2O$ in the lower half of the atmosphere. In Case 2, we observe more mixing the blue region of the predicted mixing zone compared to the other test cases, which can be attributed to compositionally dense air sinking from the upper part of the atmosphere to the lower part of the atmosphere.}
	\label{fig:case2_verticalprofiles}
\end{figure*}

\begin{figure*}[ht]
	\centering
	\includegraphics[width=\textwidth]{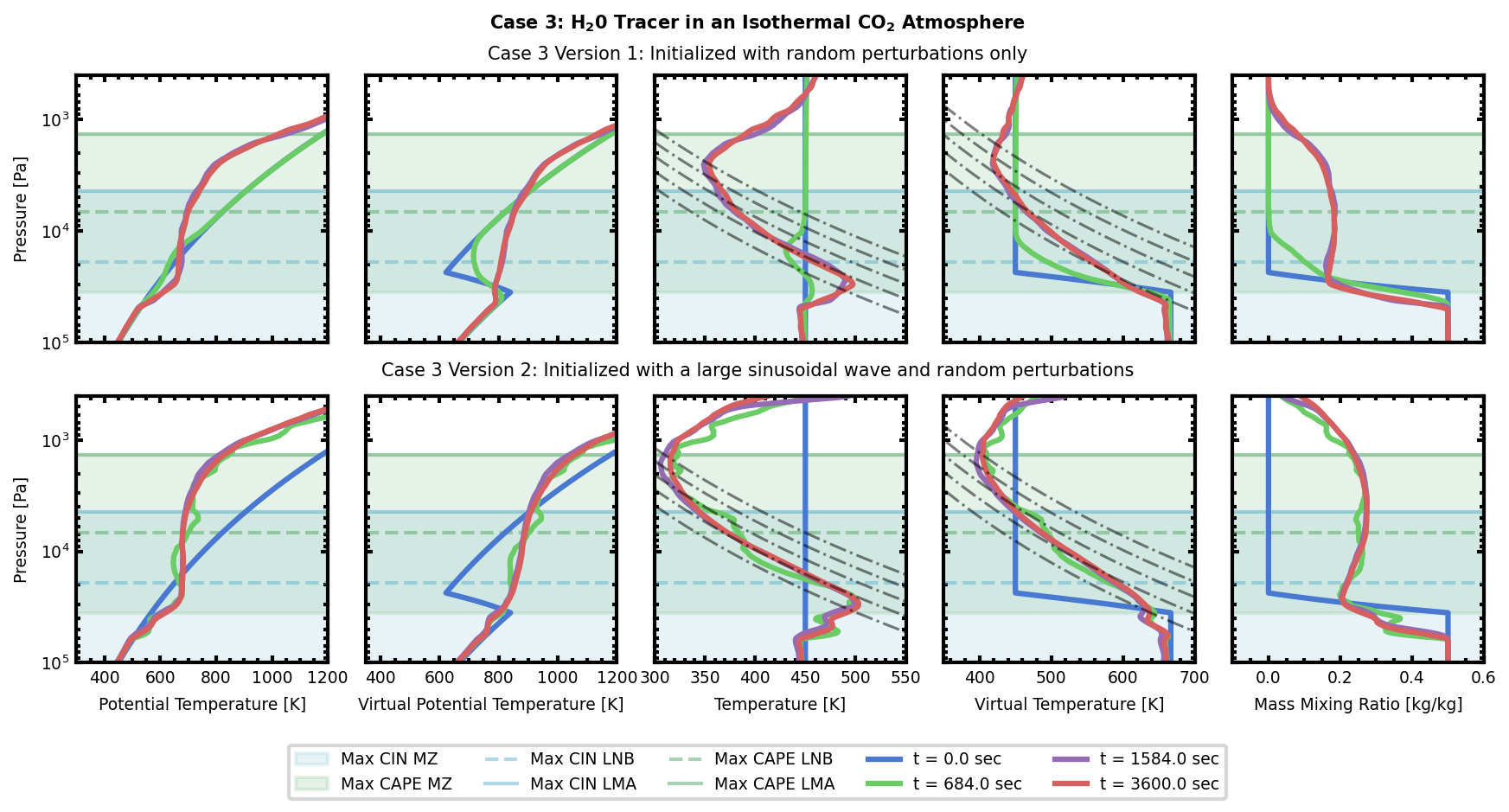}

	\caption{Same as Fig.~\ref{fig:case1_verticalprofiles} but showing the CM1 vertical profiles for for Case 3, isothermal $\rm CO_2$ atmosphere with an initial non-condensing $\rm H_2O$ in the lower half of the atmosphere. As expected, we observe greater compositional enhancement of buoyancy in the $\rm CO_2$ atmosphere compared to the Earth air atmosphere with the same initial $\rm q_v(p)$ profile shown in Fig.~\ref{fig:case1_verticalprofiles}.}
	\label{fig:case3_verticalprofiles}
\end{figure*}

\begin{figure*}[ht!]
	\centering
	\includegraphics[width=\textwidth]{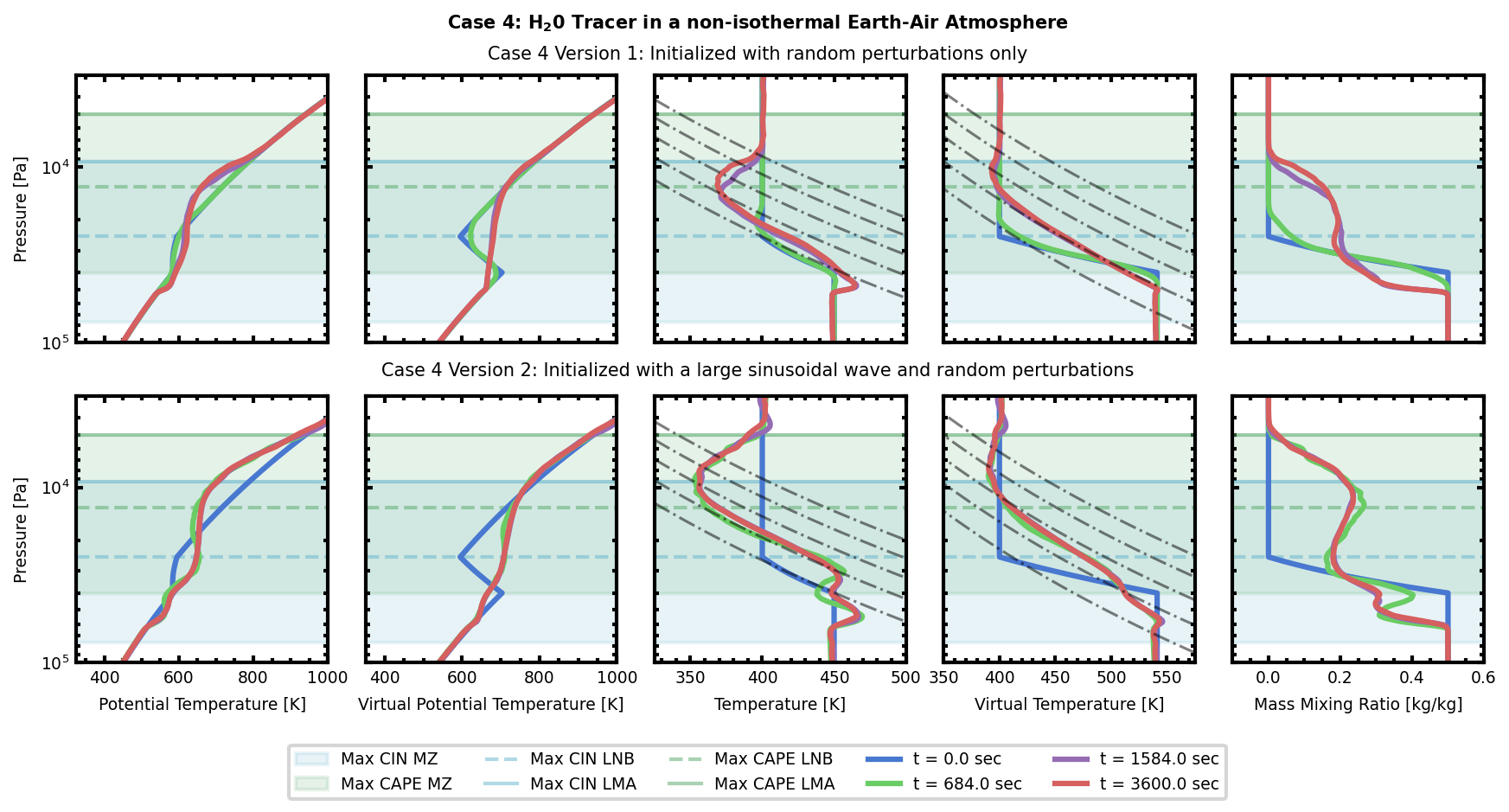}

	\caption{Same as Fig.~\ref{fig:case1_verticalprofiles} but showing the CM1 vertical profiles for for Case 4, Earth air atmosphere with an initial step profile in both $\rm T$ and $\rm q_v$ profiles. Case 4 has a larger predicted mixing zone than Case 1 due to the both temperature and composition contributing to buoyancy.}
	\label{fig:case4_verticalprofiles}
\end{figure*}

\begin{figure*}[ht]
	\centering
	\includegraphics[width=\textwidth]{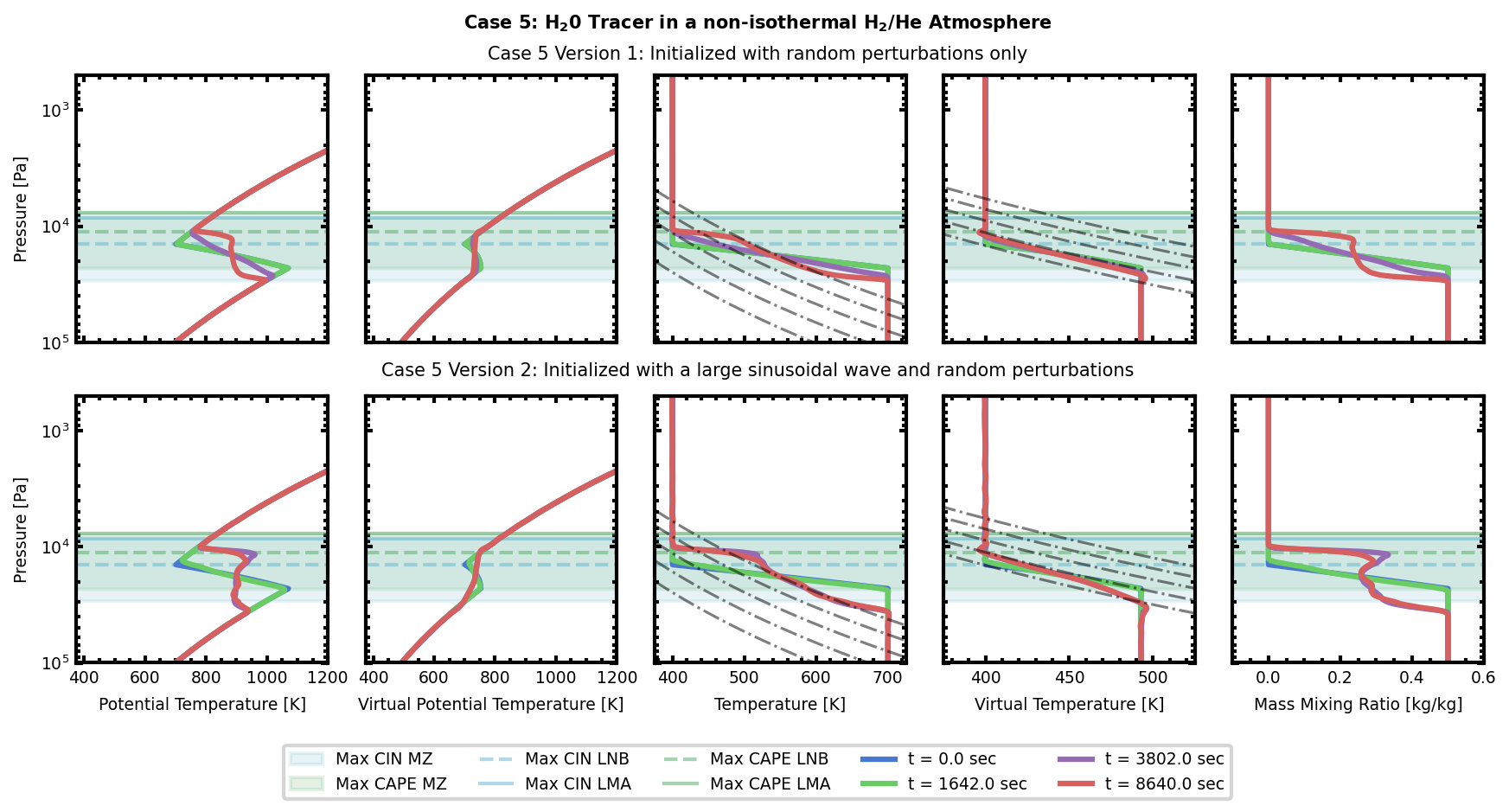}

	\caption{Same as Fig.~\ref{fig:case1_verticalprofiles} but showing the CM1 vertical profiles for for Case 5, $\rm H_2$ atmosphere with an initial step profile in both $\rm T$ and $\rm q_v$ profiles. Case 5 is the only CM1 test case performed where the compositional gradient suppresses buoyancy while the thermal gradient enhances buoyancy. Case 5 shows the least amount of convective mixing within the vertical domain, which illustrates that the compositional gradient can strongly stabilize the atmospheric profile against convection in $\rm H_2$ atmospheres.}
	\label{fig:case5_verticalprofiles}
\end{figure*}

\begin{figure*}[ht]
	\centering
	\includegraphics[width=\textwidth]{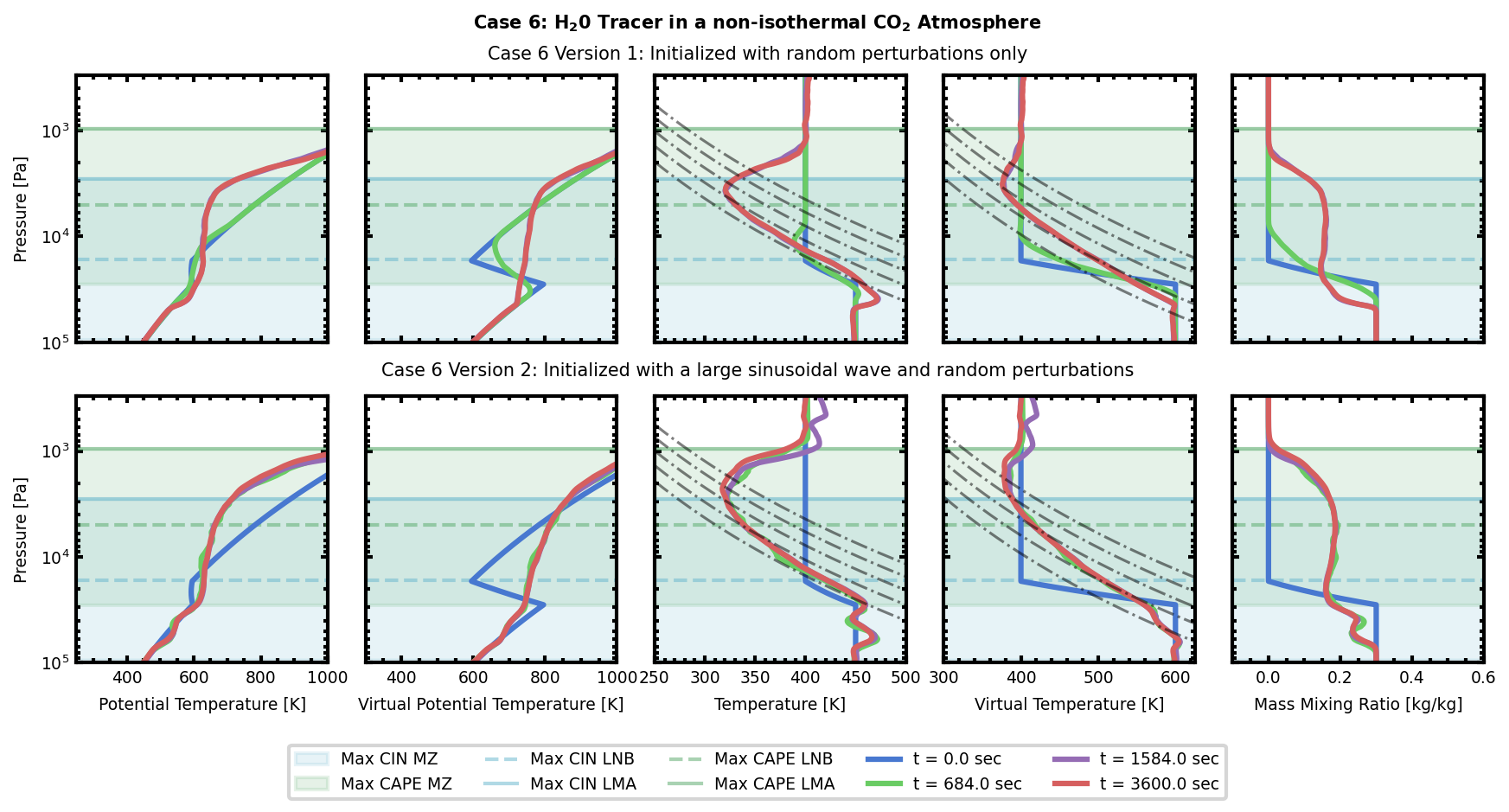}

	\caption{Same as Fig.~\ref{fig:case1_verticalprofiles} but showing the CM1 vertical profiles for for Case 6, $\rm CO_2$ atmosphere with an initial step profile in both $\rm T$ and $\rm q_v$ profiles. Case 6 used an initial value of $\rm q_v,s = 0.3$ kg/kg in the lower atmosphere due the CM1 becoming too unstable and crashing when using higher values of $\rm q_v,s$.}
	\label{fig:case6_verticalprofiles}
\end{figure*}

\subsection{High Resolution CM1 Simulation Results Overview}
Figs.~\ref{fig:air_2d_mass_mixing_ratio}--\ref{fig:co2_2d_mass_mixing_ratio} show vertical cross sections of mass mixing ratio taken in the middle of the y-domain at seven time steps throughout the CM1 simulation for all six high resolution test cases. Fig.~\ref{fig:air_2d_mass_mixing_ratio} shows the vertical cross sections for the Earth air atmosphere with Cases 1 and 4 in the top and bottom panel respectively. Fig.~\ref{fig:h2_2d_mass_mixing_ratio} shows the vertical cross sections for the $\rm H_2$ atmosphere with Case 2 in the top panel and Case 5 in the bottom panel. Fig.~\ref{fig:co2_2d_mass_mixing_ratio} shows $\rm q_v$ vertical cross sections for $\rm CO_2$ atmosphere with Case 3 in the top panel and Case 6 in the bottom panel. Within each panel, the top row of Figs.~\ref{fig:air_2d_mass_mixing_ratio}--\ref{fig:co2_2d_mass_mixing_ratio} shows the $\rm q_v$ cross sections when CM1 is initialized with initial random perturbations in $\rm \theta$, while the bottom row shows the mass mixing ratio when CM1 is run with an initial large sinusoidal perturbation in addition to random perturbations in $\rm \theta$. In Case 2 shown in the top panel of Fig.~\ref{fig:h2_2d_mass_mixing_ratio}, the yellow and black regions are flipped compared to the other test cases since $\rm H_2O$ tracer was placed in the upper part of the atmosphere to create an initial unstable compositional gradient.

The initial step profile which is the basis for our compositional convection CM1 test cases is an instance of the Rayleigh-Taylor instability. The Rayleigh-Taylor instability describes a fluid dynamic instability at the interface of two fluids where a lighter fluid is pushing a heavier fluid \citep{Zhou2017, Gauthier2010}. A visual characteristic of Rayleigh-Taylor instabilities are mushroom-shaped plumes. Rayleigh-Taylor instabilities manifest as normal mode instabilities of the system linearized about the state of rest, and grow exponentially until adjacent plumes start mixing and the instability transitions to a turbulent mixing regime \citep{Zhou2017}. The formation, growth, and mixing of mushroom-shaped clouds is clearly visible within the mass mixing ratio cross sections shown in Figs.~\ref{fig:air_2d_mass_mixing_ratio}--\ref{fig:co2_2d_mass_mixing_ratio}.

The vertical cross sections of $\rm q_v$ in Figs.~\ref{fig:air_2d_mass_mixing_ratio}--\ref{fig:co2_2d_mass_mixing_ratio} show that the large-scale initial perturbation creates a single coherent plume which rises a finite amount and then mixes horizontally within the domain. The simulations initialized with a large-scale perturbation show that the convecting plume penetrates deeper into the vertical domain creating a deeper mixing layer relative the simulations initialized with small-scale perturbations. Additionally, the mass mixing ratio does not mix as uniformly within the horizontal domain in the simulations initialized with a large-scale perturbation. Conversely, when CM1 is initialized with random perturbations, small-scale plumes form within the horizontal domain, grow and start to mix the horizontal domain and a subset of the vertical domain, but do not penetrate as deep into vertical domain compared to the simulation results when using an initial large sinusoidal perturbation.

The vertical cross sections presented in Figs.~\ref{fig:air_2d_mass_mixing_ratio}--\ref{fig:co2_2d_mass_mixing_ratio} supports our assumption in section~\ref{sec:atm_thermo_background} that an air parcel maintains its composition as it rises/sinks adiabatically within the atmosphere. It is clear in the bottom rows of Figs.~\ref{fig:air_2d_mass_mixing_ratio}--\ref{fig:co2_2d_mass_mixing_ratio} that as an air parcel is rising, it keeps its composition from the parcel's initial location. After the air parcel seems to have reached marginal stability, convection begins to mix the composition of the parcel with the environment.

Additionally, when comparing Case 1 to the Case 4 in Fig.~\ref{fig:air_2d_mass_mixing_ratio}, it is visually apparent that Case 4 mixes deeper within the vertical domain. This is because Case 4 uses sounding profile 2 in which positive buoyancy is generated by both the $\rm T$ and $\rm q_v$ profiles. In contrast, in Case 1, buoyancy is generated solely from the compositional gradient. This trend is not observed when comparing Case 3 and Case 6 in Fig.~\ref{fig:co2_2d_mass_mixing_ratio} because we reduced the $\rm q_v,s = 0.3$ in Case 6 due to CM1 becoming too numerically unstable. If we had used $\rm q_{v,s} = 0.5$ for the initial state in Case 6, we would expect convective mixing deeper in the atmospheric column compared to Case 3. The bottom panel of Fig.~\ref{fig:h2_2d_mass_mixing_ratio} shows Case 5, where we had a large temperature gradient overcome the compositional suppression of buoyancy. Case 5 shows the least amount of mixing of the $\rm H_2O$ tracer within the atmospheric column which can be attributed to the competing effects of the temperature and compositional gradient. In all six high resolution test cases, the $\rm H_2O$ tracer does not fully homogenize the vertical column but is seen to form stable layers of varying composition.

We evaluated the CAPE of the initial atmospheric sounding state, using Eqn.~\ref{eq:cape}, for all of our high resolution CM1 test cases to determine the predicted mixing zone -- the region in which we expect convection to occur. The predicted mixing zone is the region within the atmosphere where an air parcel, if perturbed a finite amount, can generate positive CAPE. The predicted mixing zone represents the region in the initial atmospheric state that is convectively unstable. To determine the predicted mixing zone, we evaluate the CAPE of air parcels starting from various heights within the initial virtual temperature profile. For the given initial atmospheric sounding state, an air parcel starting at $\rm p_1$ (the pressure at the start of the smeared out density discontinuity) generates the maximum amount of positive CAPE, which sets the upper bound for the predicted mixing zone. Any parcel starting $\rm p_1~>~p~>~p_2$, will generate positive CAPE when lifted a finite amount but produces less CAPE than than the parcel that is lifted from $\rm p_1$. Any parcel starting from $\rm p~<~p_2$ will only generate negative CAPE and is convectively stable. An air parcel starting from $\rm p_1~<~p~<~p_{min \; CAPE}$ will first generate negative CAPE, but if the parcel is lifted enough, at some point the rising air parcel's adiabat will intersect the atmospheric virtual temperature profile. At this point, the air parcel has reached an unstable neutral buoyancy point, and if the parcel is lifted upward an infinitesimal amount the parcel will freely rise in the atmosphere until it reaches its LNB. $\rm p_{min \; CAPE}$ is the pressure location greater than $\rm p_1$, where an air parcel's virtual adiabat would intersect at $\rm p_2$. An air parcel rising from $\rm p \ge p_{min \; CAPE}$ will only generate negative CAPE and is convectively stable. $\rm p_{min \; CAPE}$ represents the lower bound in the atmospheric column from where an air parcel can no longer generate positive CAPE. We set the lower bound of the predicted mixing zone at $\rm p_{min \; CAPE}$. In some cases $\rm p_{min \; CAPE}$ was calculated to be greater than $\rm p_s$, so we adjusted $\rm p_{min \; CAPE}$ = $\rm p_s$. The lower region of the predicted mixing zone represents the maximum amount of negative CAPE a parcel needs to overcome to freely rise within the initial sounding state.

In Figs.~\ref{fig:case1_verticalprofiles}-\ref{fig:case6_verticalprofiles}, we show the predicted mixing zone as the shaded blue-green region. The green shaded region represents the upper bound of the mixing zone and is determined by a parcel starting at $\rm p_1$, while the blue shaded region is the lower bound of the predicted mixing zone given by a parcel rising from $\rm p_{min \; CAPE}$. White regions represent convectively stable regions. Both the blue and green regions start at their respective initial parcel locations and end at their respective LMA. The dark green region is the overlap of the blue and light green regions. Within each panel of Figs.~\ref{fig:case1_verticalprofiles}--\ref{fig:case6_verticalprofiles} we show the LNB as a dashed horizontal line and LMA as a solid horizontal line for the both the upper and lower bound of the predicted mixing zone in green and blue respectively.

Figs.~\ref{fig:case1_verticalprofiles}--\ref{fig:case6_verticalprofiles} show the vertical atmospheric sounding state for each of the high resolution CM1 test cases respectively. In each figure, the top row shows the final atmospheric state of the CM1 test case initialized with small-scale random perturbations in $\rm \theta$. The bottom row shows the final atmospheric state for the same CM1 test case but initially perturbed in $\rm \theta$ with both random perturbations and a sinusoidal wave given by Eqn.~\ref{eq:3d_wave}. The five panels in each row show (from left to right) the vertical atmospheric profile of potential temperature, virtual potential temperature, temperature, virtual temperature, and mass mixing ratio at four different time steps from the CM1 simulation as shown by different coloured solid lines. The vertical profiles for temperature, potential temperature, and mass mixing ratio are calculated by taking the horizontal average of the output 3D CM1 data for $\rm T$, $\rm \theta$, and $\rm q_v$ as a function of height. The virtual temperature and virtual potential temperature profiles are calculated by using the horizontally averaged 1D temperature and mass mixing ratio profiles with the equations defined in section~\ref{sec:atm_thermo_background}. The black dashed lines plot uniform composition dry adiabats in the temperature panel, calculated using Eqn.~\ref{eq:DryAdiabat}, where $\rm \beta_{mix}$ is constant and has a value of $\rm R_b / C_{p,b}$. The black dashed lines plot virtual adiabats, calculated using Eqn.~\ref{eq:VirtualAdiabatExact}, in the virtual temperature profile panel, where $\rm \beta_{mix}$ is assumed to be an approximately constant value given by $\rm \beta_{mix}(p_1)$. The effect of using different values for $\rm \beta_{mix}$ in calculating the virtual adiabat is relatively minor. We present animations of the full 3D evolution of mass mixing ratio for all six of our high resolution CM1 simulations in the Appendix.

\subsection{Discussion of CM1 Results}
\subsubsection{What is the final compositional profile after convective mixing in an initially inhomogeneous composition atmosphere?} \label{sec:final_qv_profile}

Within Figs.~\ref{fig:case1_verticalprofiles}--\ref{fig:case6_verticalprofiles}, the mass mixing ratio panel shows that $\rm q_v$ does not fully homogenize within the predicted mixing zone since a stable atmospheric state was reached before this could occur. Mixing is observed to occur within a subset of the predicted mixing zone for all six high resolution test cases, creating a staircase-like profile for the final state mass mixing ratio. Case 3 version 2 shown by the bottom panel in Fig.~\ref{fig:case3_verticalprofiles} is observed to have the largest amount of mixing and nearly homogenizes the final state mass mixing ratio. When using the large sinusoidal perturbation (version 2 results) in Case 4 shown in Fig.~\ref{fig:case4_verticalprofiles}, and Case 6 shown in Fig.~\ref{fig:case6_verticalprofiles}, there seem to be two ``stairs" of $\rm q_v$ that form within the predicted mixing zone. The mass mixing ratio profile's ``compositional staircases" are similar to results of compositional convection effects in stellar atmospheres and thermohaline convection within Earth's oceans \citep{Garaud2015, Radko2003}. The formation of compositional staircases suggest that there are discrete layers of varying values of $\rm q_v$ that form within the final state. This is also seen in the 2D vertical cross sections of $\rm q_v$ shown in Figs.~\ref{fig:air_2d_mass_mixing_ratio}--\ref{fig:co2_2d_mass_mixing_ratio} for all six high resolution test cases. Note, the formation of compositional staircases poses a challenge for accurate parameterization of compositional convection, because the shape of the final $\rm q_v$ profile becomes hard to predict without numerical simulations.

\begin{figure*}[ht]
	\centering
	\includegraphics[width=\textwidth]{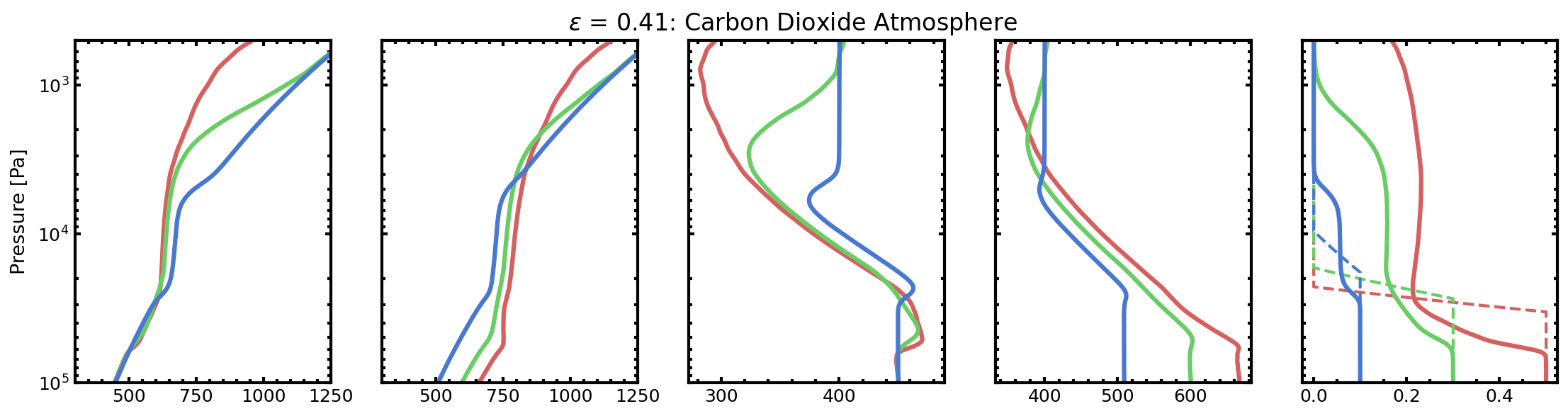}
	\includegraphics[width=\textwidth]{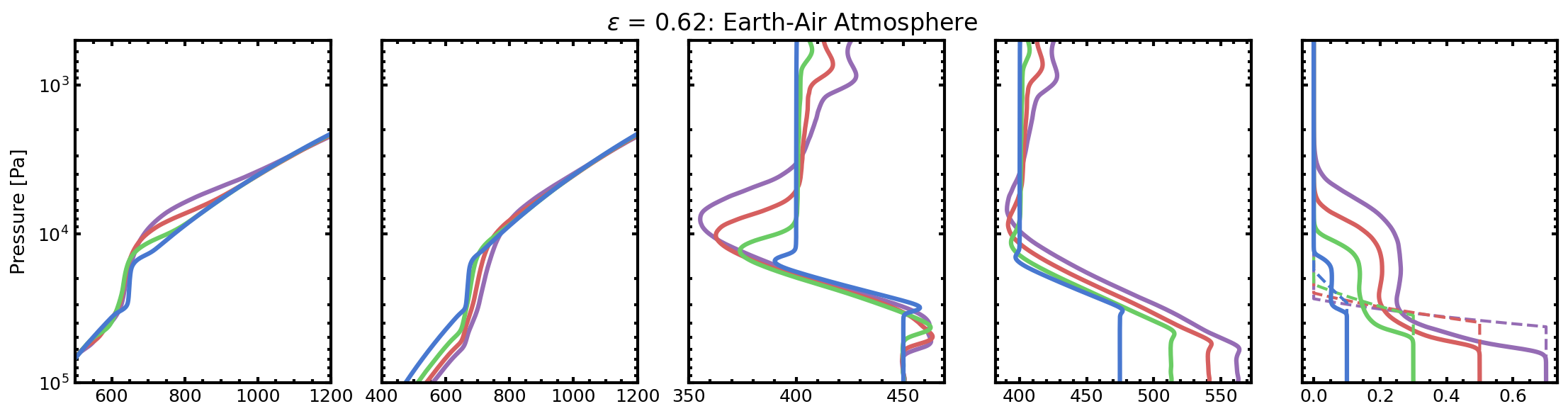}
	\includegraphics[width=\textwidth]{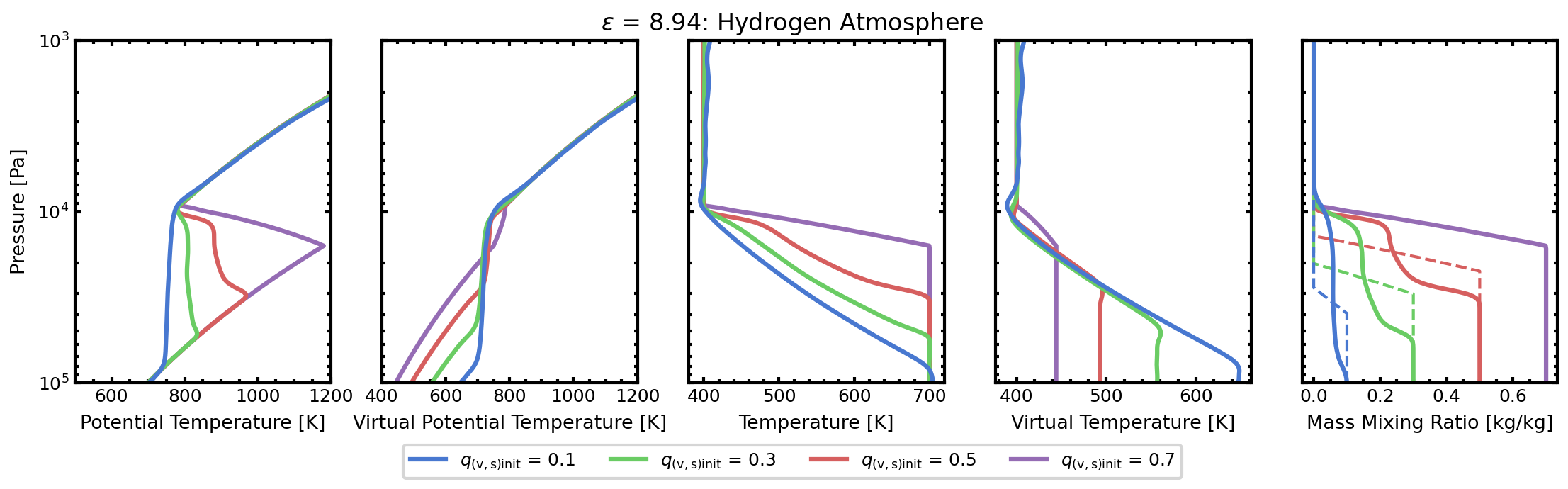}

	\caption{Parameter study of the effect of varying the initial compositional gradient while keeping the initial temperature profile fixed. The initial atmospheric state for these simulations is defined by sounding profile 2 with: 1. temperature step profile with $\rm T_{upper}$ = 400 K and $\rm T_{lower}$ = 450 K for $\rm \epsilon < 1$ cases, and $\rm T_{lower}$ = 700 K for $\rm \epsilon > 1$ case; 2. an initial $\rm q_v$ step profile with $\rm q_{v, \; upper}$ = 0.0 and $\rm q_{v, \; lower}$ varying from 0.1 to 0.7. The dashed profiles in the mass mixing ratio panel plot the initial atmospheric state $\rm q_v$ profiles for reference. For the $\rm \epsilon < 1$ cases, increasing the compositional gradient leads to more vigorous convective mixing and for the $\rm CO_2$ atmosphere, more homogenization of the final state mass mixing ratio profile. Conversely, for the $\rm \epsilon > 1$ case, increasing the compositional gradient leads to the suppression of convection. The vertical profiles presented here are averages of the horizontal domain. All simulations presented here are run for 8640 seconds. The resolution was reduced to save computation time.}

	\label{fig:qvstudy}
\end{figure*}

\begin{figure*}[ht]
	\centering
	\includegraphics[width=\textwidth]{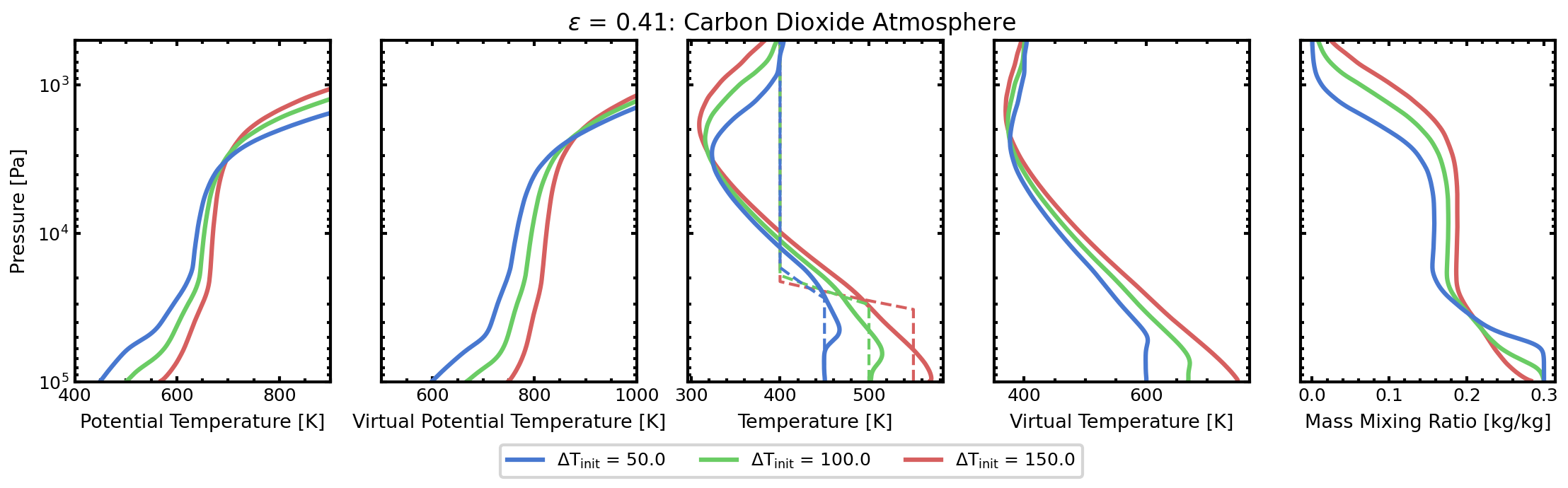}
	\includegraphics[width=\textwidth]{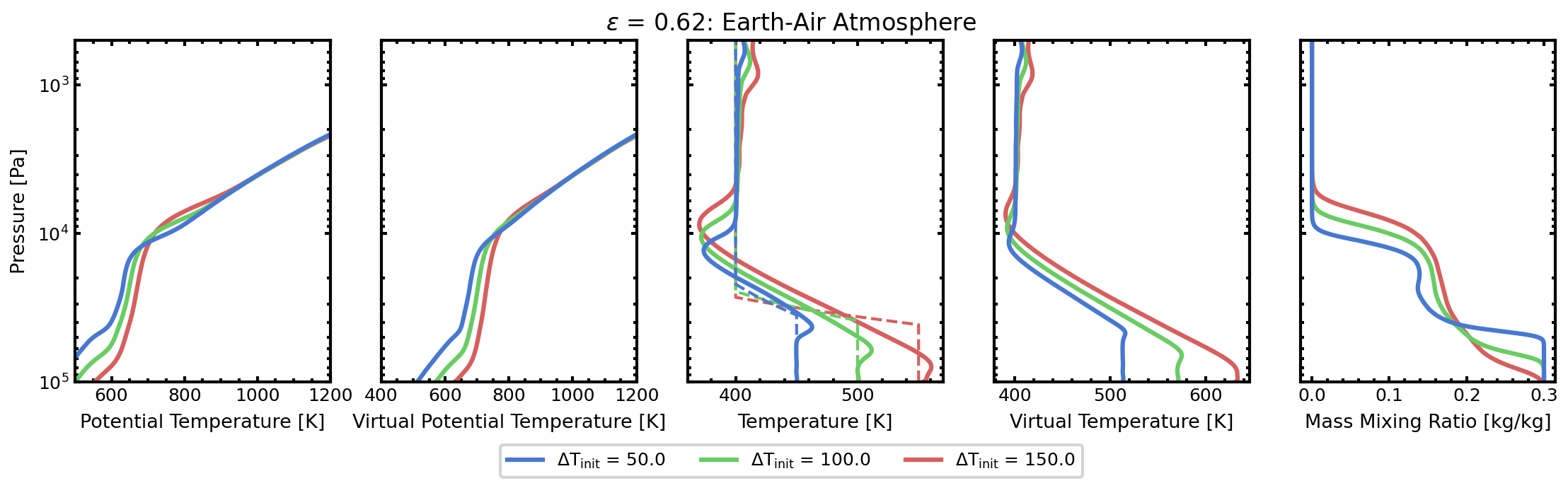}
	\includegraphics[width=\textwidth]{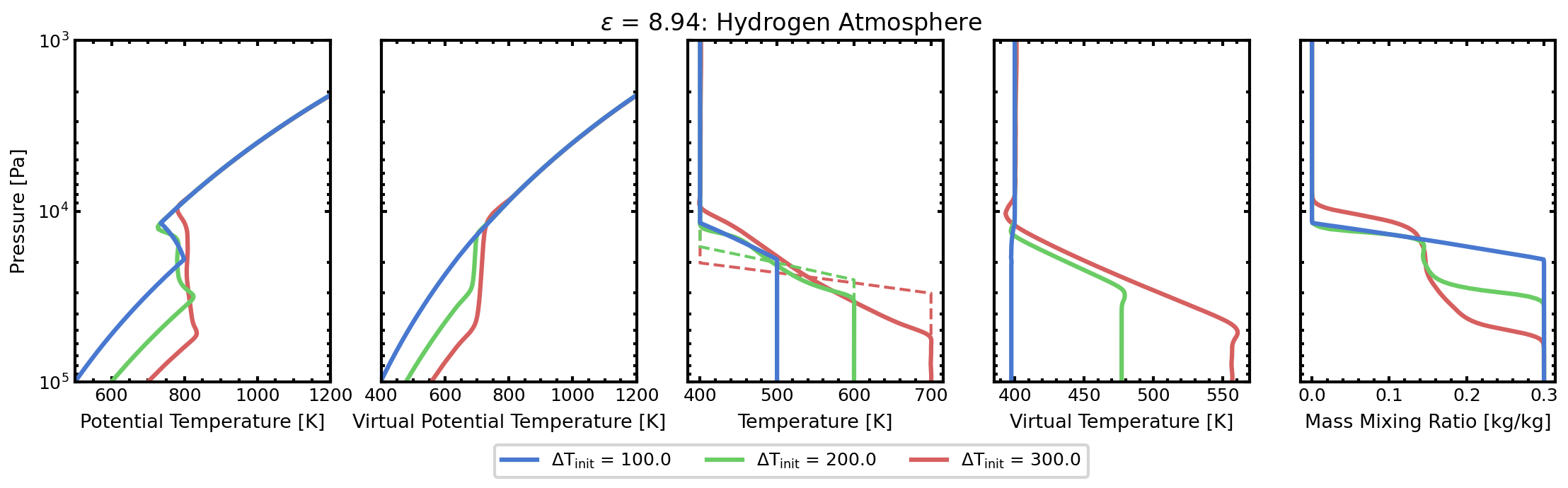}

	\caption{Parameter study of the effect of varying the initial temperature gradient while keeping the initial compositional profile fixed. The initial atmospheric state for these simulations is defined by sounding profile 2 with: 1. temperature step profile with $\rm T_{upper}$ = 400 K and $\rm T_{lower}$ varying from 450 K to 550 K for $\rm \epsilon < 1$, and $\rm T_{lower}$ varying from 500 K to 700 K for $\rm \epsilon > 1$ case; 2. an initial $\rm q_v$ step profile with $\rm q_{v, \; upper}$ = 0.0 and $\rm q_{v, \; lower}$ = 0.3. The dashed profiles in the temperature panel plot the initial atmospheric state profiles for reference. $\rm \Delta T_{init}$ = $\rm T_{lower}$ - $\rm T_{upper}$ for the initial state atmosphere. For the $\rm \epsilon < 1$ cases, increasing the temperature gradient leads to more vigorous convective mixing and more homogenization of the final state mass mixing ratio profile. For the $\rm \epsilon > 1$ case, increasing the temperature gradient leads to the formation of larger compositional staircases. The vertical profiles presented here represent averages of the horizontal domain. All simulations presented here are run for 8640 seconds. The resolution was reduced to save computation time.}

	\label{fig:tstudy}
\end{figure*}

To further explore how convection mixes the atmospheric tracer in initially inhomogeneous composition atmospheres, we performed two parametric studies with sounding profile 2 where we (1) varied the initial lower atmospheric mass mixing ratio value, $\rm q_{v, s, initial}$, while keeping the temperature profile constant shown in Fig.~\ref{fig:qvstudy}; (2) varied the initial lower atmosphere $\rm T_s$ value while keeping the $\rm q_{v}$ profile fixed shown in Fig.~\ref{fig:tstudy}. Varying the lower atmospheric mass mixing ratio and temperature leads to varying the compositional and temperature gradient respectively for our CM1 simulation setup.

Fig.~\ref{fig:qvstudy} shows the final state atmospheric profiles for CM1 simulations starting with $\rm q_{v, s, initial}$ ranging from 0.1 to 0.7 while keeping the temperature profile constant. In the cases where $\rm \epsilon < 1$, shown in the top ($\rm CO_2$ atmosphere) and middle row (Earth Air atmosphere) of Fig.~\ref{fig:qvstudy}, increasing the initial amount of atmospheric tracer in the lower atmosphere further enhances convective buoyancy and results in more mixing. The $\rm CO_2$ atmosphere case experiences a significantly larger compositional enhancement of buoyancy compared to the Earth Air atmosphere case. Due to this, having a larger initial $\rm q_{v, s, initial}$ in the $\rm CO_2$ atmosphere leads to more homogenization of the final state compositional profile relative to the Earth air atmosphere.

In the case where $\rm \epsilon > 1$ illustrated in the bottom row of Fig.~\ref{fig:qvstudy}, increasing the compositional gradient leads to less vertical mixing and eventually, the compositional gradient completely shuts off convection. When $\rm q_{v, s, initial}$ = 0.1, convection mixes the final state lower atmosphere area to be uniform because the temperature gradient is able to overcome the compositional suppression of buoyancy. However, when 0.1 $\rm > q_{v, s, initial} > $ 0.7, we see the clear formation of compositional staircases. At $\rm q_{v, s, initial}$ = 0.7, we observe no convection because the compositional gradient completely stabilizes the temperature gradient. In this case, the initial atmospheric state is Ledoux stable, $\rm T_{v, parcel} < T_{v, amb}$ everywhere, therefore no convection was observed within CM1. Note, that in the $\rm \epsilon > 1$ case we used a temperature step profile with $\rm \Delta T$ between the upper and lower atmosphere of 300 K compared to the $\rm \epsilon < 1$ case where $\rm \Delta T = $ 50 K. A large temperature step is needed in the $\rm H_2$ atmosphere case to overcome the strong stabilization of convection due to the compositional gradient. We do not observe any convection in the $\rm H_2$ atmosphere case when $\rm \Delta T = $ 50 K, even for the $\rm q_{v, s, initial}$ = 0.1 case, because the initial sounding state is Ledoux stable.

Similarly, we show the effect of varying the initial lower atmosphere $\rm T_s$ while keeping the initial mass mixing ratio profile fixed with $\rm q_{v, s, initial}$ = 0.3 and $\rm q_{v, 1, initial}$ = 0.0 using sounding profile 2 in Fig.~\ref{fig:tstudy}. When $\rm \epsilon < 1$, shown in the top ($\rm CO_2$ atmosphere) and middle rows (Earth Air atmosphere), increasing the temperature gradient leads to more vigorous convective mixing. When $\rm \epsilon < 1$, increasing the initial temperature gradient causes the final atmospheric $\rm q_v$ to transition from a compositional staircase like regime to a more well-mixed homogeneous state. When $\rm \epsilon > 1$, shown in the bottom row of Fig.~\ref{fig:tstudy}, increasing the temperature gradient does increase the convective mixing but at a much slower rate relative to the $\rm \epsilon < 1$ cases. In all three simulations run in the $\rm \epsilon > 1$ case, the final mass mixing ratio profiles still show the formation of compositional staircases.

The parameter study in Figs.~\ref{fig:qvstudy}~-~\ref{fig:tstudy} is presented to highlight that the final state after convective mixing in an initially inhomogeneous composition atmosphere can still have a compositional gradient. For a large portion of the parameter study, we observe the formation of compositional staircases. The formation of compositional staircases suggests that a marginally stable final atmospheric state is reached prior to convection being able to fully homogenize the mixing layer (the region which was was initially unstable to convection). Again, the formation of compositional staircases poses a challenge to accurately parameterizing compositional convection.

\subsubsection{Does convection mix the final atmospheric state of an initially inhomogeneous composition atmosphere to a local or global neutrally stable state?}
Convection relaxes to a state of global neutral stability when $\rm q_v$ homogenizes within the convectively unstable layer, i.e. the predicted mixing zone. When the final atmospheric state has compositional inhomogeneities within the mixing layer, the most general marginal stability condition is Ledoux local marginal stability (see section~\ref{sec:atm_thermo_background}). A large portion of the CM1 simulation results presented here show the formation of compositional staircases and variation of $\rm q_v$ within the predicted mixing zone suggesting that convection will likely mix an initially inhomogeneous composition atmosphere to a final \textit{local} marginally stable state.

In Figs.~\ref{fig:case1_verticalprofiles}--\ref{fig:case6_verticalprofiles}, convective mixing observed within the simulations is limited to a sub-portion of the vertical domain that was initially unstable (blue-green shaded region) and is seldom observed to penetrate into regions that are initially statically stable. Regions outside of the area where convective mixing primarily occurred kept their initial statically stable profiles. We only observe the convective plumes penetrating into statically stable regions in Case 1 and 3 when using using the large-scale initial perturbation (version 2). In both of these test cases, the convective plume penetrated into the statically stable region before mixing with the horizontal domain. Note, accurately modelling penetrating convective plumes is a challenge in developing a convection parameterization scheme.

The virtual potential temperature profiles in Figs.~\ref{fig:case1_verticalprofiles}--\ref{fig:case6_verticalprofiles}, mimic a staircase like structure as well, suggesting the final state forms stably stratified layers. The same trend is observed in $\rm \theta(p)$ for all cases except Case 5. In Case 5, shown in Fig.~\ref{fig:case5_verticalprofiles}, the potential temperature profile shows a region where $\rm d\theta/dz~<~0$, however, the corresponding $\rm d\theta_v/dz$ is approximately constant suggesting that the compositional profile is still stabilizing the temperature profile in the final atmospheric state.

For the atmosphere to mix to a global marginally stable state, additional energy would need to be supplied to the system to further mix the atmosphere. Case 3 and 6 are the two cases that nearly reach global stability. Both of these cases have a large amount of CAPE in the initial atmospheric sounding, and also have the largest predicted mixing zones. Within the parameter study, the cases that are seen to nearly homogenize are cases where the density difference between the upper and lower atmosphere is the largest, and the system has substantially more CAPE in the initial state. Within the parameter study, again the $\rm CO_2$ atmosphere test cases are the ones which are observed to mix towards homogenizing the atmosphere. We do not observe the Earth Air atmosphere cases mixing toward a global marginally stable state. This is because the $\rm CO_2$ atmosphere has a significantly larger compositional enhancement of buoyancy compared to the Earth Air atmosphere.

CAPE can be a good indicator for determining weather the atmospheric system has enough energy to mix to a state of global or local marginal stability. Convection will mix an initially inhomogeneous composition atmosphere to a final state of local stability unless the initial state of the atmosphere has enough CAPE to fully mix the convectively unstable regions.

\subsubsection{If convection mixes the atmosphere to a locally stable state, can it be described by the virtual adaibat?}

The virtual temperature profiles shown in Figs.~\ref{fig:case1_verticalprofiles}--\ref{fig:case6_verticalprofiles} evolve towards a virtual adiabat. When using small-scale initial displacements (top row of Figs.~\ref{fig:case1_verticalprofiles}--\ref{fig:case6_verticalprofiles}), the virtual temperature profile clearly evolves to have a final state that lies on a virtual adiabat. When using the large-scale initial perturbation, (bottom row of Figs.~\ref{fig:case1_verticalprofiles}--\ref{fig:case6_verticalprofiles}), the trend of the final atmospheric state evolving towards a virtual adaibat is still present but less clear. When using the large-scale initial perturbation, we observed in Figs.~\ref{fig:air_2d_mass_mixing_ratio}-\ref{fig:co2_2d_mass_mixing_ratio} that the horizontal domain does not mix as uniformly as when only using small-scale perturbations. Some of the deviation of the final $\rm T_v(p)$ state from the virtual adiabat in the bottom row of Figs.~\ref{fig:case1_verticalprofiles}--\ref{fig:case6_verticalprofiles} can be attributed to the horizontal domain variation since these profiles are calculated as horizontal averages. In Case 1 version 2, shown in Fig.~\ref{fig:case1_verticalprofiles}, convection started at a $\rm p~<~p_1$, and the virtual adiabats plotted in Fig.~\ref{fig:case1_verticalprofiles} are calculated for a parcel rising from $\rm p_1$.

The final state observed within the CM1 simulations is a virtual adiabat which describes a marginally Ledoux stable state (see section~\ref{sec:atm_thermo_background}). The final atmospheric state is a marginally stable state that depends on both the final temperature and mass mixing ratio profiles. Without compositional variation in the final atmospheric state, the uniform composition dry adiabat specifies a unique state for the marginally stable system. When there is compositional variation in the final atmospheric state, there is a whole range of marginally stable final temperature and compositional profiles that correspond to the same virtual adiabat. Our CM1 simulations show that one of the possible marginally stable compositional profiles is realized instead of the case were $\rm q_v$ completely homogenizes in the predicted mixing zone. This implies that a convection parameterization scheme needs to account for how convection redistributes composition.

Figs.~\ref{fig:case1_verticalprofiles}-\ref{fig:case4_verticalprofiles}, and Fig.~\ref{fig:case6_verticalprofiles} show the final $\rm T(p)$ profile does not fully lie on the uniform composition dry adiabat but rather increases in the lower part of the atmosphere and cools in the upper part of the atmosphere. The temperature evolution can be attributed to compositionally buoyant air parcels rising from the lower part of the atmosphere and cooling the atmosphere adiabatically as the air parcels ascend and vice versa. The extra compositional buoyancy allowed plumes to rise further than they otherwise would. For case 5 shown in Fig.~\ref{fig:case5_verticalprofiles}, the temperature decreases in the lower part of the predicted mixing zone and heats in the upper part of the atmosphere. Case 5 is the only case where we have compositional suppression of buoyancy and the compositional buoyancy is opposing the thermal buoyancy. When the atmosphere mixes in Case 5, convection expends energy to lift a heavy air parcel from the lower part of the atmosphere causing the temperature to decrease in the lower part of the predicted mixing zone. As a dense air parcel sinks, energy is released and the upper region of the predicted mixing zone heats. For all CM1 test cases studied here, convection transports energy upwards which is seen by the virtual temperature profile heating the upper part of the mixing region while cooling in the lower part of the mixing zone.

Our simulation results for Case 5 show a trend in the final atmospheric temperature profile that is similar to the findings from compositional convection simulations presented in \cite{DaleyYates2021}. In their study, \cite{DaleyYates2021} used a 3D hydrodynamics code to simulate a hot, rocky exoplanet atmosphere with a chemical reaction (i.e. forcing term) in the upper atmosphere that converted CO to $\rm CO_2$, thus maintaining a compositional gradient within the atmosphere. The \cite{DaleyYates2021} simulation setup used an initial stabilizing temperature gradient with an initial destabilizing compositional gradient, i.e. compositional and thermal buoyancy are opposing each other. Even though the details of the simulation setup in \cite{DaleyYates2021} differ from that presented in this work, their final temperature profile shows the same trend we observe for Case 5 where the lower atmospheric temperature deceases and the upper atmospheric temperature increases.

\subsubsection{How do large-scale versus small-scale perturbations affect the final state after convective mixing?}

Comparing results from the top and bottom rows of Figs.~\ref{fig:case1_verticalprofiles}--\ref{fig:case6_verticalprofiles} show that using different initial perturbations results in notable differences in the final atmospheric state predicted by CM1. When using only random perturbations (seen in the top row of Figs.~\ref{fig:case1_verticalprofiles}--\ref{fig:case6_verticalprofiles}) convection starts at the discontinuity location $\rm p_1$, and mixing occurs primarily in a subset of the predicted mixing zone given by the green region of the predicted mixing zone which is from [$\rm p_1$, LMA maximum CAPE limit]. When using a large-scale sinusoidal perturbation (bottom row of Figs.~\ref{fig:case1_verticalprofiles}--\ref{fig:case6_verticalprofiles}), convective mixing is observed to occur in a larger extent of the predicted mixing zone. In particular, there is more observed mixing within the lower part of the blue shaded region and the upper part of the green maximum CAPE region. In some cases, the system ``over-stabilizes," i.e. leaves the virtual adiabat more stable than it needs to be, when using the large-scale perturbation. This is most pronounced in Cases 1, 3 and 6. Accounting for over-stabilization is an additional challenge to consider when developing a convection parameterization scheme.

For test case 1 version 2 in Fig.~\ref{fig:case1_verticalprofiles}, and case 3 version 2 in Fig.~\ref{fig:case3_verticalprofiles} there is observed mixing outside of the predicted mixing zone. We believe the mixing outside of the predicted mixing zone bounds is due to convective overshoot, and is not enough to homogenize the atmospheric tracer or adjust to a virtual adiabat. Overall, when using a large-scale initial perturbation, an unstable air parcel; 1. can penetrate deeper in the atmosphere without mixing due to small-scale turbulent process, 2. accesses more CAPE which leads to greater mixing than a parcel which is only displaced a small amount. Using the large-scale initial perturbation suppresses the faster-growing small-scale instabilities which equilibrate non-linearly at smaller amplitudes, with less penetration.

In general, the CM1 simulations show that the final state is sensitive to the scale of the initial atmospheric perturbation that triggers convection. The vertical extent within the predicted mixing zone where convection is observed to occur within all six high resolution test cases, varies based on the initial scale of the perturbation. This is difficult to represent within a convection parameterization scheme, because a priori, the scale of the atmospheric perturbation is unknown.

\section{Non-Condensing Convection Adjustment Scheme} \label{sec:adj_scheme}

\begin{figure*}[ht]
	\centering
	\includegraphics[width=0.49\textwidth]{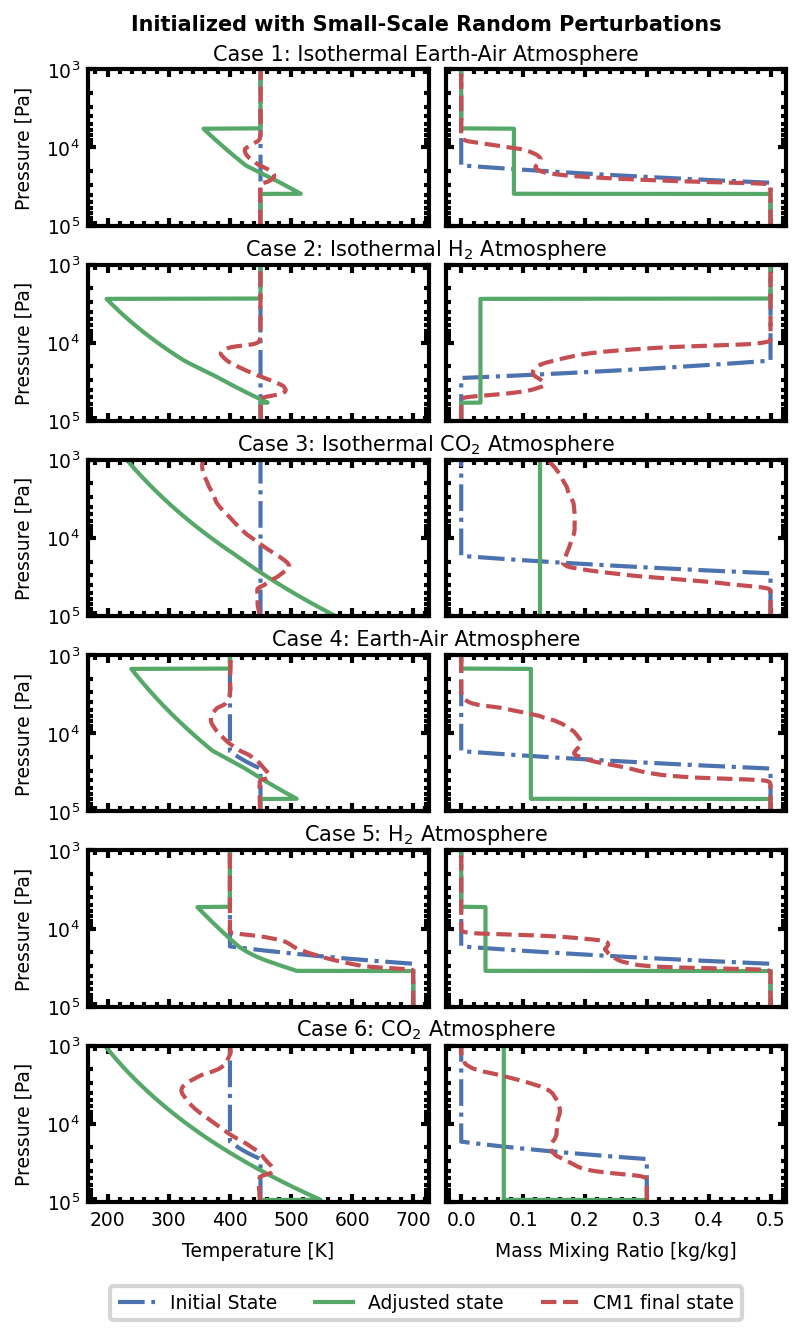}
 \includegraphics[width=0.50\textwidth]{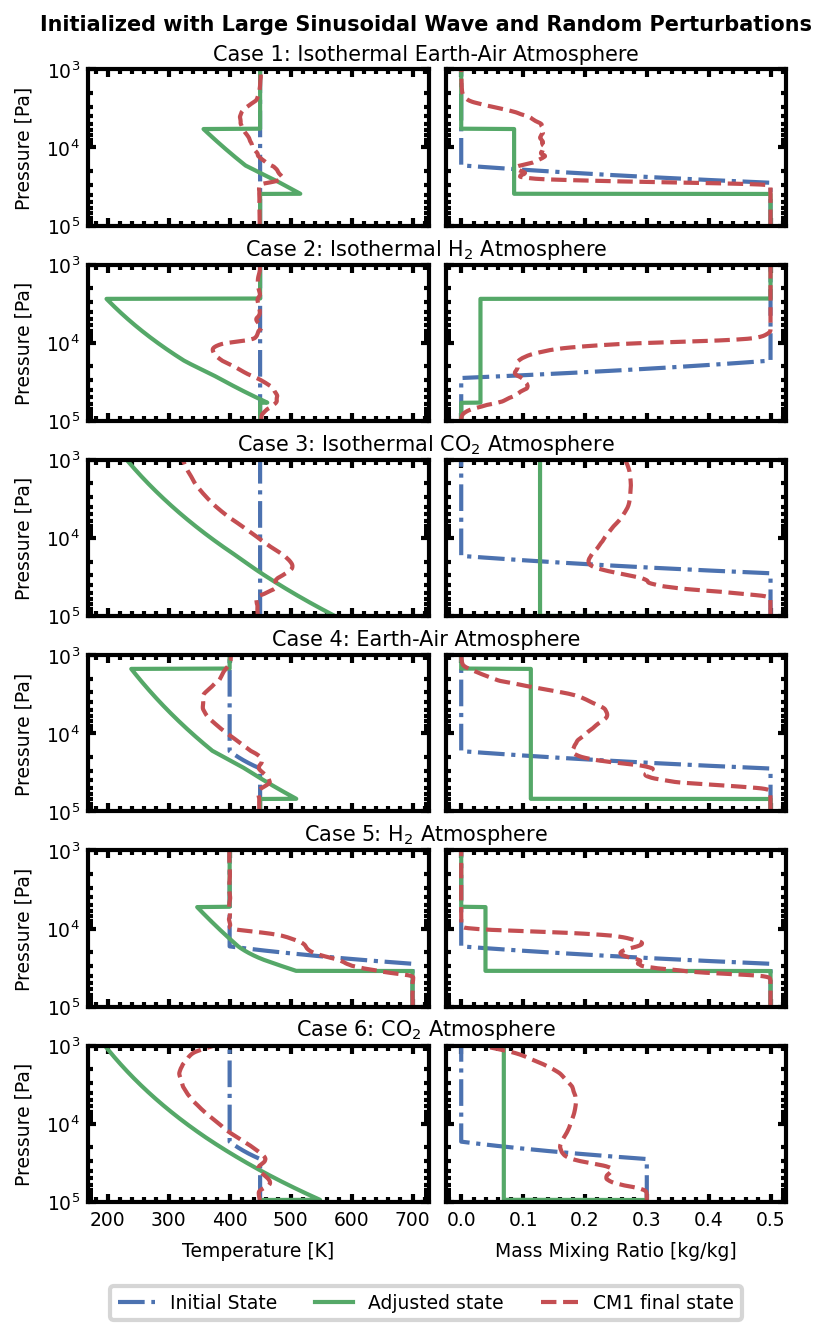}
	\caption{Comparison of the non-condensing convective adjustment scheme to CM1 simulation results for CM1 Cases 1-6. The blue dash-dotted line plots the initial atmospheric state, the red dashed line plots the final atmospheric state from the CM1 simulation, and the green solid line shows the predicted final state using our non-condensing convective adjustment scheme. The left panel shows the comparison of the CM1 results to the convective adjustment scheme for the CM1 simulation run using only random perturbations to trigger convection, while the right panel shows the results when convection is started using random perturbations and a large sinusoidal wave given by Eqn.~\ref{eq:3d_wave}. We observe relatively good agreement between the adjustment scheme and CM1 results. A noticeable discrepancy is that the adjustment scheme is unable to represent convective overshoot and has sharp cut-offs for the adjusted atmospheric profile due to using the predicted mixing zone as hard bounds.}
	\label{fig:adjustment_scheme}
\end{figure*}

One of the main findings from our CM1 modelling results is that the final atmospheric state, after convective mixing in initially inhomogeneous composition atmospheres, is most generally described by a marginally stable state that is defined by the virtual adiabat. We use this result as the starting point to develop a non-condensing convective adjustment scheme that can account for compositional variation. Our convective adjustment scheme (1) limits convection only to occur within the predicted mixing zone and; (2) adjusts the final atmospheric state to lie on virtual adiabat.

Our convective adjustment scheme uses a CAPE analysis of the initial atmospheric state to determine the region in which convection occurs. We use the same CAPE analysis performed previously to create the blue-green shaded region to determine the bounds in which we expect convection to occur. We set the lower convection bound to match the lower bound of the blue shaded region: the lowest point within the atmosphere from which a parcel can rise to create positive CAPE. We set the upper bound of the convection region to be the LMA of the maximum CAPE region, which corresponds to the green shaded region in Figs.~\ref{fig:case1_verticalprofiles}--\ref{fig:case6_verticalprofiles}. The upper and lower bounds were chosen to be inclusive of all areas in which convection is observed within our CM1 simulations. This does mean that the mixing layer within the convective adjustment scheme can be larger than what was actually observed in the CM1 simulation results, especially for the simulation test cases that used a small-scale initial perturbation.

The virtual adiabat depends on both the final temperature and mass mixing ratio profiles. There is a whole family of temperature and mass mixing ratio profiles that can create the same virtual adiabat. Therefore, to create a convective adjustment scheme that adjusts the initial atmospheric state to a virtual adiabat, we need to a priori make an assumption regarding either the final state temperature or mass mixing ratio profile. Adjusting to a virtual adiabat is a degenerate problem because $\rm T_v$ is dependant on both T and $\rm q_v$. For the convective adjustment scheme presented in this work, we make an assumption regarding the final state mass mixing ratio. We assume that the final state mass mixing ratio will uniformly mix within the region where convection occurs allowing us to adjust the temperature profile to lie on a virtual adiabat. Simply, we take the pressure-weighted mean of the mixing ratio within the the predicted mixing zone as determined by the CAPE analysis. This assumption is imperfect and not fully representative of the CM1 simulation results, but it provides an initial framework to incorporate compositional buoyancy effects in convection parameterizations.

It is worth noting that the final state atmospheric profiles are dependent on the initial compositional profile in two ways when assuming $\rm q_v$ homogenizes in the predicted mixing zone. First, the initial mass mixing ratio profile impacts the CAPE analysis used to define the boundaries of convection for our convection parameterization scheme. CAPE is calculated using the $\rm T_v(p)_{initial}$ profile which is a function of $\rm q_v(p)_{initial}$ and $\rm T(p)_{initial}$. Second, since the convective adjustment scheme only homogenizes the final state mass mixing ratio within the predicted mixing zone, any region outside the CAPE analysis will retain its original value from the initial state $\rm q_v$ profile. This approach allows us to be able to create a compositional staircase like structure within the final state and maintain a compositional gradient within $\rm q_v(p)_{final}$. The impact of the initial mass mixing ratio profile on the final $\rm q_v$ state is more pronounced when the predicted mixing zone occupies a smaller portion of the vertical domain (i.e. when there is compositional stabilization of convection). However, due to the assumption we make regarding the final $\rm q_v$ profile, our ``compositional stairs" will be inaccurate compared to the CM1 simulation results.

Only in the case where the predicted mixing zone encompasses the entire vertical domain does the final state atmospheric profiles become independent of the initial compositional gradient in the convective adjustment scheme. In this particular case, the convection adjustment scheme predicts the final state to be well mixed everywhere, resulting in a fully uniform final $\rm q_v$ profile. The virtual adiabat for a well-mixed atmosphere becomes equivalent to the dry adiabat with the appropriate $\rm \beta_{mix}$ value for the well mixed atmosphere. In this particular case, the final state profile is solely dependent on the total amount of the initial tracer.

The output for our convective adjustment scheme is shown in Fig.~\ref{fig:adjustment_scheme}. We applied our convective adjustment scheme on the same initial atmospheric sounding profiles that we used to run the CM1 test cases 1 -- 6. In Fig.~\ref{fig:adjustment_scheme}, the left panel shows the temperature and mass mixing ratio for the 6 high resolution test cases explored in this work initialized with small-scale perturbations. The right panel shows $\rm q_v$ and $\rm T$ profiles for all 6 high resolution test cases initialized with the large-scale sinusoidal perturbation. In each plot, the initial atmospheric state is shown by the blue dash-dotted line, the CM1 final state is shown by the red dashed line, and the final state predicted by the convective adjustment scheme is shown by the green solid line. Overall, the convective adjustment scheme is able to match the CM1 outputs decently. The adjustment scheme does predict sharp cut-offs in the vertical profiles due to the hard bound of convection only occurring within the predicted mixing zone from the CAPE analysis. There is some discrepancy between the final state predicted from the adjustment scheme compared to our CM1 outputs that we attribute to our assumption of determining the convection bounds and the $\rm q_v$ of the final adjusted state. For the simulations that used small initial random perturbations in $\rm \theta$ to trigger convection, a better assumption for the upper bound of convection would have been the LNB of the maximum CAPE region (green shaded region), since convection was not observed above this point in those simulations. However, when using a large-scale initial perturbation to trigger convection, we do observe mixing in some simulations between the LNB and LMA for the maximum CAPE region. Since we cannot predict the scale of the initial disturbance that will trigger convection, to be more inclusive we use the LMA of the maximum CAPE region as the upper bound of the predicted mixing zone for our convective adjustment scheme.

Assuming $\rm q_v$ becomes uniform within the mixing zone is a decent first order approximation. However, this approximation implies that the $\rm q_v$ profile variation within the predicted mixing zone, and multiple steps seen in Case 4 and Case 6 are not captured by the convective adjustment scheme. Additionally, we are not able to match the final $\rm q_v$ value in the mixing zone for Case 2 and Cases 4 - 6. Fig.~\ref{fig:adjustment_scheme} shows that the convection bounds used for these cases are too large compared to the region in which convection was observed within the CM1 simulations. By assuming a homogeneous $\rm q_v$ within the predicted mixing zone, we are susceptible to under-predicting the value of $\rm q_v$ in the convective adjustment scheme. Despite the inaccuracy, the assumption of a well-mixed end state within the mixing zone still captures the key features of the end state temperature profile in most cases. Additionally, our convectively adjustment scheme while energetically consistent and yielding a stable final state, is unable to produce compositional staircases within the predicted mixing zone, account for convective overshoot or penetrative convection. The CM1 simulations highlight these phenomena are important, and future convection parameterization schemes need to address these shortcomings. Our convective adjustment scheme is imperfect, but provides a starting point to represent non-condensing compositional convection in GCMs and is better than ignoring compositional buoyancy altogether for sub-Neptune and super-Earth atmospheres.

\section{Conclusions} \label{sec:conclusion}
We performed initial value problem simulations of compositional convection within three different planetary atmospheres including an Earth-air, $\rm CO_2$, and $\rm H_2$ atmosphere. The high resolution CM1 simulations modelled a convectively unstable initial atmosphere with two initial atmospheric states and two initial perturbations to trigger convection. We also performed two parameter study simulation sets to explore how sensitive the final atmospheric state is to the initial compositional and thermal gradient respectively. The main finding of our CM1 modelling results are:
\begin{enumerate}
    \item The final atmospheric compositional profile shows the formation of compositional staircases in a large parameter space. The formation of compositional staircases is more frequently observed than convection homogenizing the mass mixing ratio within the entire mixing layer.

    \item Within the convectively mixed region, the final atmospheric state is a locally marginally stable state, satisfying marginal stability according to the Ledoux criterion.

    \item The final atmospheric virtual temperature profile is well approximated by a virtual adiabat in the mixing layer, where we assume $\rm \beta_{mix}$ is a representative constant for the region mixed by convection. However, the temperature profile itself is sensitive to the presence of compositional staircases remaining in the final state.

    \item The CM1 simulations results show sensitivity to large-scale vs small-scale convection triggers. Mixing is observed to occur over a larger extent of the vertical domain for simulations that initially trigger convection using a large-scale perturbation. The predicted mixing zone captures the primary region in which convection occurs, but always overestimates the region in which mixing occurs when convection is triggered using small-scale initial perturbations. The predicted mixing zone was observed to overestimates the region in which mixing occurred when using the large-scale perturbation for the $\rm H_2$ atmosphere case, which has strong compositional stabilization of buoyancy.

    \item Development of a convective adjustment scheme is challenging because it must be able to account for;
    \begin{enumerate}
        \item The formation of compositional staircases in the final $\rm q_v$ profile in many circumstances.
        \item Sensitivity to large-scale versus small-scale convection triggers
    \end{enumerate}
\end{enumerate}

Based on the CM1 simulations, we developed a non-condensing convective adjustment scheme that relaxes an initial atmospheric state to a final state defined by the virtual adiabat. Adjusting to a local, marginally stable state described by the virtual adiabat is a degenerate problem. In order to adjust the final atmospheric state to a virtual adiabat, a priori knowledge or an assumption regarding either the final compositional or temperature profile is required. We compared our convective adjustment scheme to our CM1 simulation outputs and found decent agreement. However, the convective adjustment scheme is currently not able to account for all the observed phenomena from the CM1 simulations. Based on observed features in the CM1 simulation results, future convection parameterization schemes that account for compositional convection should be able to:
\begin{enumerate}
    \item Predict the compositional staircases in the final mass mixing ratio profile
    \item Better represent the bounds of convection (i.e. mixing layer) based on the scale of the initial atmospheric perturbation
    \item Account for convective overshoot
    \item Account for cases which may have convective penetration.
\end{enumerate}

Future work of modelling compositional convection needs to incorporate the effects of condensation and radiative forcing. We still need to test how expectations from our initial value problem simulations align with statistical equilibrium situations influenced by radiative heating and cooling. Additionally, further work is needed to improve representation of compositional convection in convection parameterization schemes.

The CM1 simulations performed in this work are aimed at improving atmospheric modelling for sub-Neptunes and super-Earths. In particular, the Earth-air and $\rm CO_2$ atmosphere cases are relevant for terrestrial planets with an ocean in the non-dilute regime, e.g. hot atmospheres which are not steam dominated. The $\rm H_2$ atmosphere cases are important to better understand sub-Neptune atmospheres, but also planets with a primordial $\rm H_2$ atmosphere which have recently formed and have not yet lost their primordial atmospheres \citep{Young2023earth}. There is still a lot of work needed to be done to fully understand these planetary atmospheres and how compositional convection might impact observations of these worlds. In particular, future work will need to incorporate condensation of water vapor and other constituents.

Convection can potentially influence exoplanet observations. Convection could result in significant water vapour injections into the upper atmosphere leading to observable cloud events \citep[see][]{Li2015}, or enable observable chemical mixing events by dredging up molecules from the deep atmosphere \citep[see][]{tsai2021inferring}. \cite{Li2015} suggested that the giant storm observed in Saturn's atmosphere in 2011 is the outcome of the interior atmosphere heating sufficiently to overcome the compositional convective inhibition caused by water vapor present within Saturn's lower atmosphere. Additionally, \cite{misener2022importance} proposed that convective inhibition due to silicate vapors in the $\rm H_2$ atmospheres of magma ocean worlds can affect the atmospheric structure of these worlds. \cite{misener2022importance} argued that the atmospheric layers created by convective inhibition make the planetary radius smaller than it would be if the planet had a purely convective, $\rm H_2$ atmosphere. To better evaluate the potential impacts of compositional convection on exoplanet observables, it is essential to first model and comprehend the behavior of convection.

\begin{acknowledgments}
N.Habib would like to thank Maxence Lefèvre and Xianyu Tan for their guidance with CM1 and helpful discussions in developing this work. This research has received support from the European Research Council (ERC) under the European Union’s Horizon 2020 research and innovation programme (Grant agreement No. 740963 to EXOCONDENSE), and from the Alfred P. Sloan Foundation under grant G202114194 to the AETHER project.
\end{acknowledgments}

\bibliography{references}
\bibliographystyle{aasjournal}

\appendix
\section{3D Animations of the CM1 Simulations}
\begin{figure}
\begin{interactive}{animation}{case1.mp4}
    \includegraphics[width=0.80\textwidth]{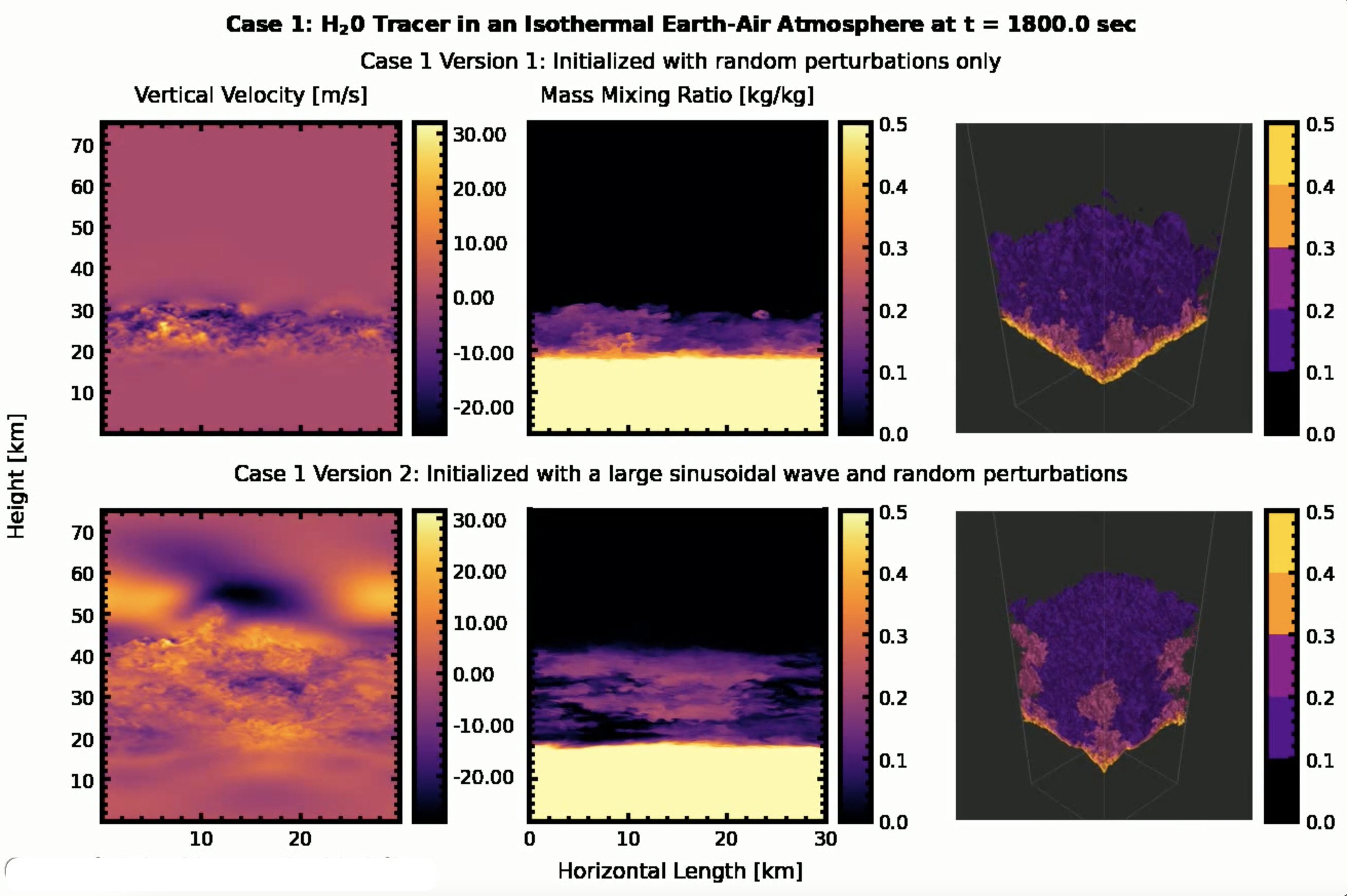}
\end{interactive}
\caption{Case 1: Earth air atmosphere with an initial isothermal temperature profile. This figure shows vertical cross sections of vertical velocity (left column) and water vapour mass mixing ratio (middle column) taken in the middle of the y-domain, and the 3D view of the mass mixing ratio (left). The top row shows the CM1 simulation results when convection is initialized with small-scale random perturbations only. The bottom row shows the CM1 simulation results when convection is triggered with both random perturbations and a large sinusoidal wave. This animation shows the formation, growth, and turbulent mixing of convective plumes due to compositional convection. The animation duration is 33 seconds. The still figure presented for the print version shows the convective mixing state at the midpoint of the animation.}
\label{anim:case1}
\end{figure}

\begin{figure}
\begin{interactive}{animation}{case2.mp4}
    \includegraphics[width=0.80\textwidth]{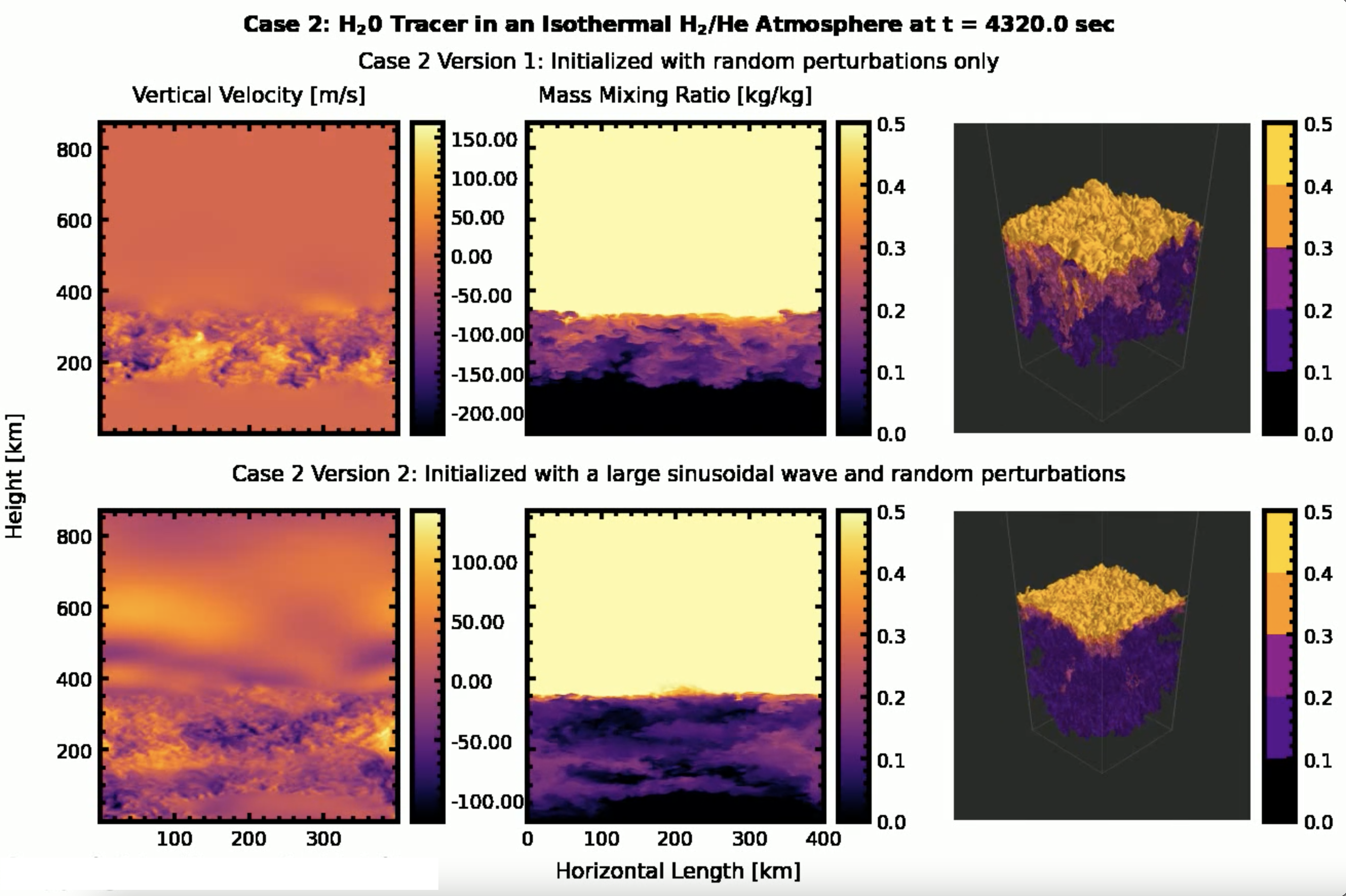}
\end{interactive}
\caption{Case 2: Hydrogen atmosphere with an initial isothermal temperature profile. Same as figure 1 but for compositional convection in a $\rm H_2$ atmosphere with an initial isothermal temperature profile.}
\label{anim:case2}
\end{figure}

\begin{figure}
\begin{interactive}{animation}{case3.mp4}
    \includegraphics[width=0.80\textwidth]{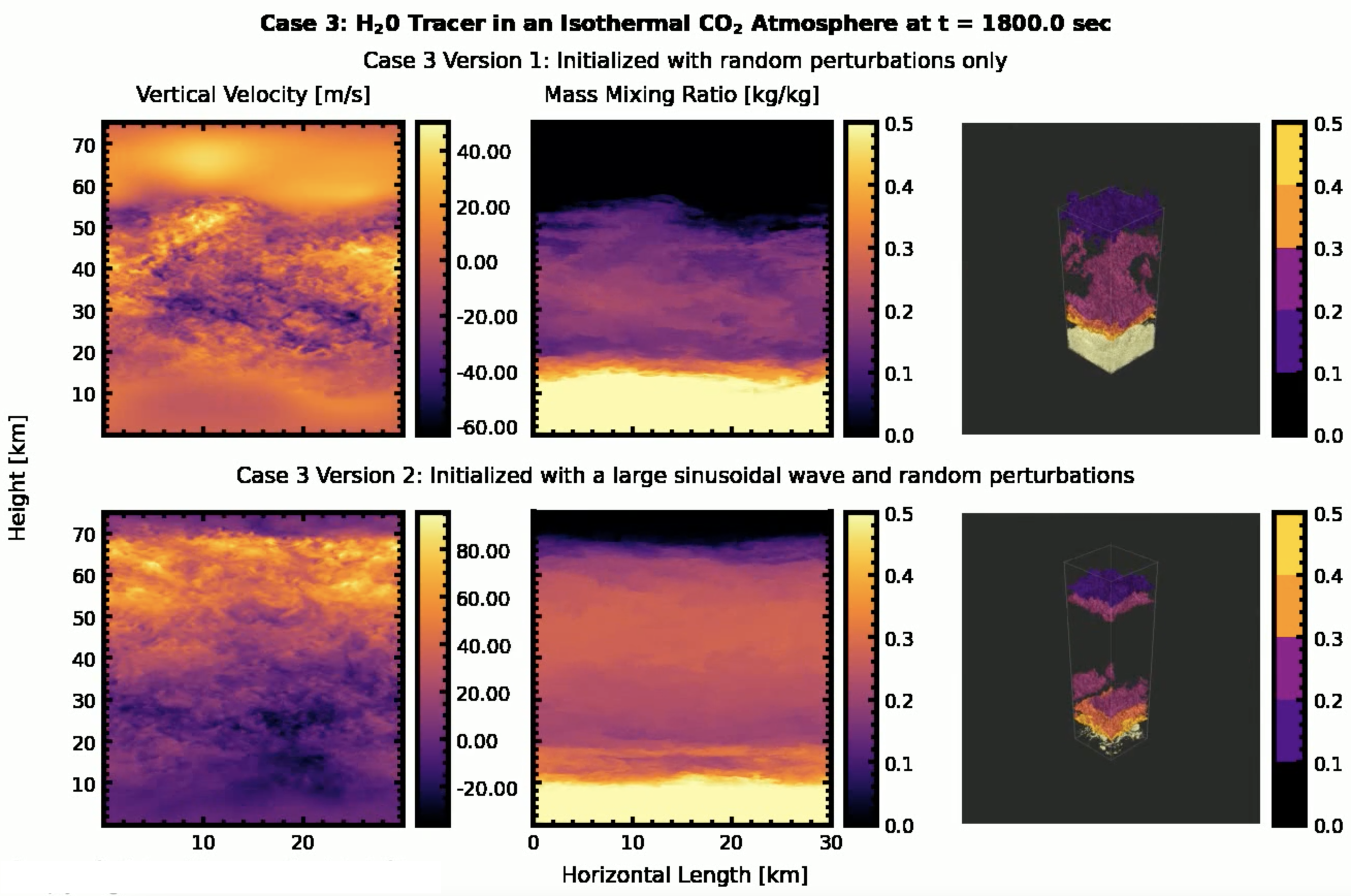}
\end{interactive}
\caption{Case 3: Carbon dioxide atmosphere with an initial isothermal temperature profile. Same as figure 1 but for compositional convection in a $\rm CO_2$ atmosphere with an initial isothermal temperature profile.}
\label{anim:case3}
\end{figure}

\begin{figure}
\begin{interactive}{animation}{case4.mp4}
    \includegraphics[width=0.80\textwidth]{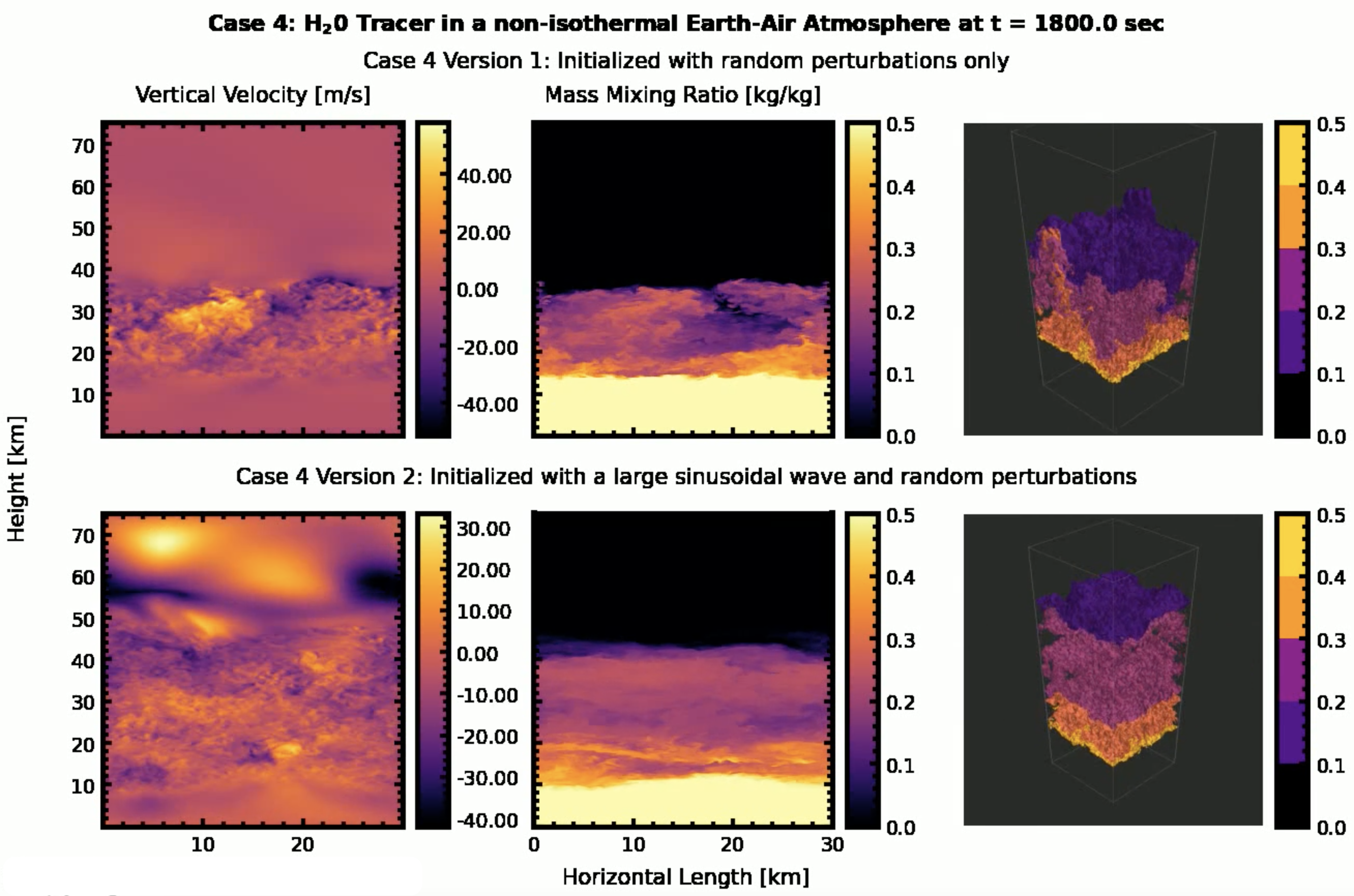}
\end{interactive}
\caption{Case 4: Earth-Air atmosphere with an initial step temperature and mass-mixing ratio profile. Same as figure 1 but for compositional convection in an Earth air atmosphere with an initial mass mixing ratio and temperature step profile.}
\label{anim:case4}
\end{figure}

\begin{figure}
\begin{interactive}{animation}{case5.mp4}
    \includegraphics[width=0.80\textwidth]{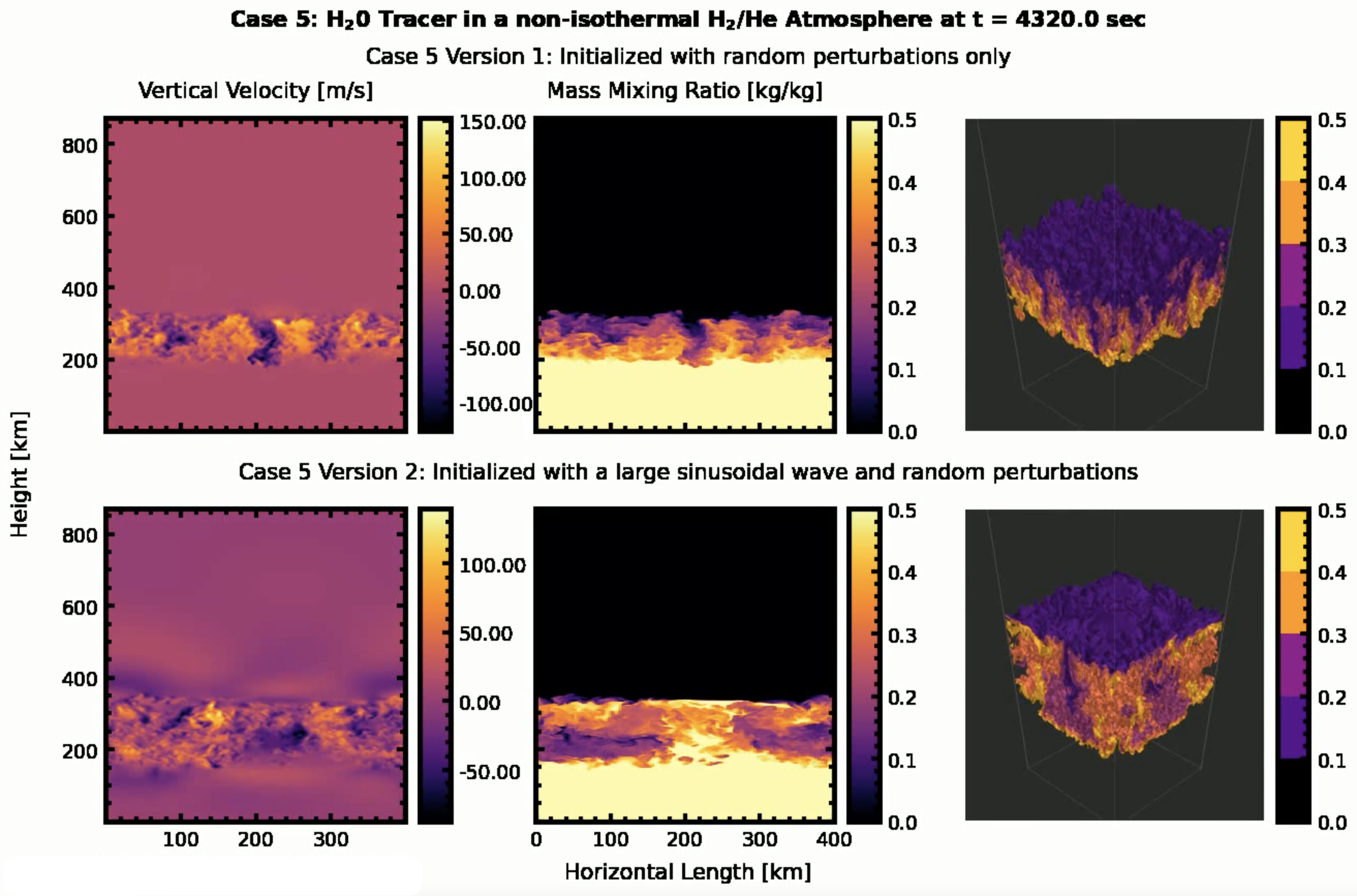}
\end{interactive}
\caption{Case 5: Hydrogen atmosphere with an initial step temperature and mass-mixing ratio profile. Same as figure 1 but for compositional convection in a $\rm H_2$ atmosphere with an initial mass mixing ratio and temperature step profile.}
\label{anim:case5}
\end{figure}

\begin{figure}
\begin{interactive}{animation}{case6.mp4}
    \includegraphics[width=0.80\textwidth]{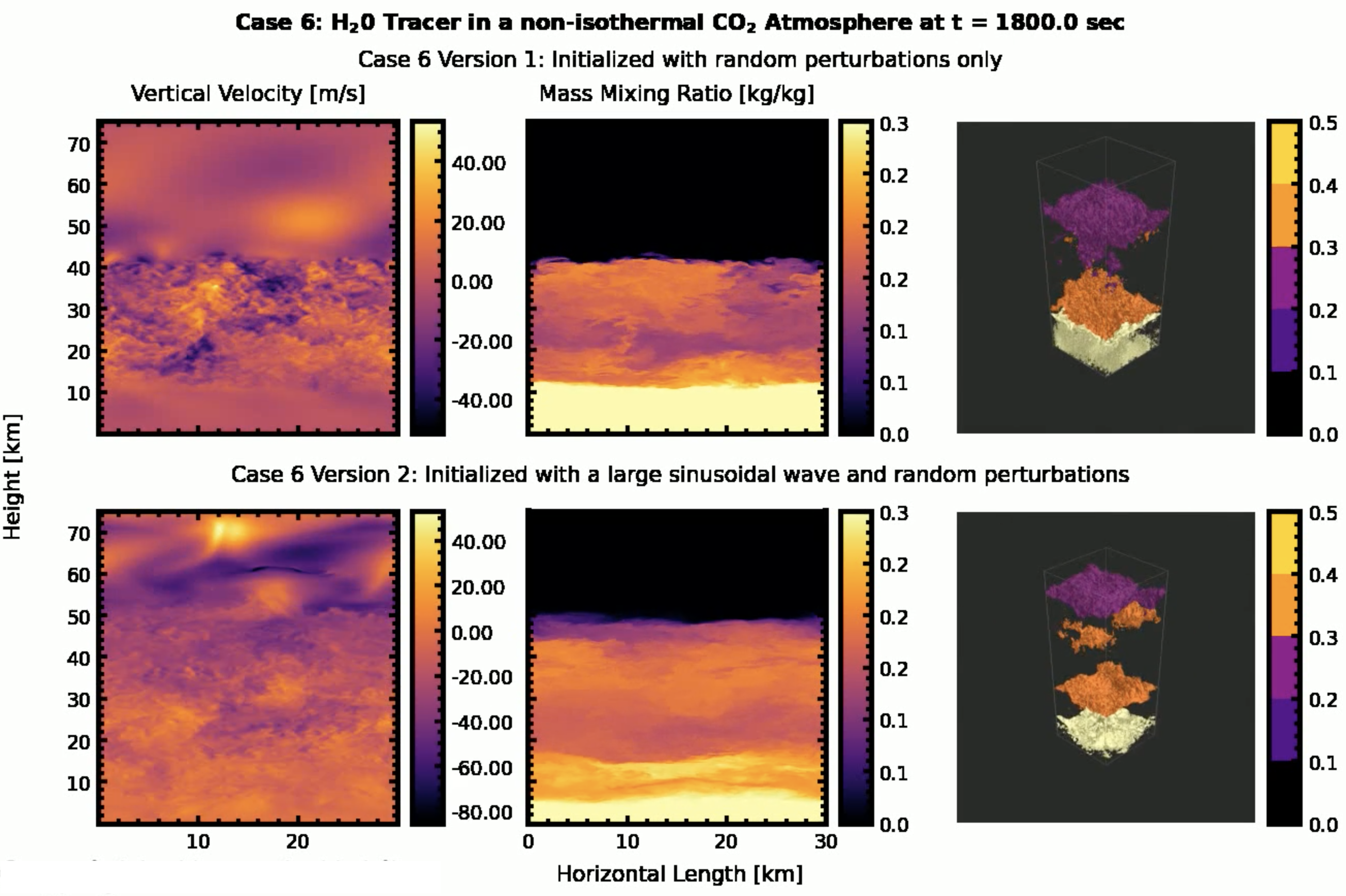}
\end{interactive}
\caption{Case 6: Carbon dioxide atmosphere with an initial step temperature and mass-mixing ratio profile. Same as figure 1 but for compositional convection in a $\rm CO_2$ atmosphere with an initial mass mixing ratio and temperature step profile.}
\label{anim:case6}
\end{figure}

We show animations from all six of our CM1 simulations in Figs.~\ref{anim:case1}--\ref{anim:case6} respectively. The top row within each figure shows the simulations results when CM1 is originally initialized with small-scale random perturbations only. The bottom row within each figure shows the simulation results when CM1 is initialized with small-scale random perturbations and a large-scale sinusoidal wave given by Eqn.~\ref{eq:3d_wave}. The left most panel shows vertical cross sections of the vertical velocity taken in the middle of the y-domain. The middle panel shows vertical cross sections of the $\rm H_2O$ mass mixing ratio taken in the middle of the y-domain. The rightmost panel shows the mass mixing ratio in 3D. The animations show the vertical velocity and mass mixing ratio as they evolve with respect to time within the CM1 simulation. For the print version, we show the still figure at the half time point of the CM1 simulation. The animations are provided to visualize, in 3D, the compositional buoyancy simulations discussed within this work.

\end{document}